\documentclass[journal]{IEEEtran}
\usepackage{cite}
\usepackage{graphicx}
\usepackage{amsmath}
\usepackage{amssymb}
\usepackage{ntheorem}
\usepackage{bm}
\usepackage{url}
\usepackage{algorithm,algorithmic}
\usepackage{color}

\newtheorem{myclm}{Property}

\newcommand{\LRT}[2]{
	\mathrel{\mathop\gtrless\limits^{#1}_{#2}}
}

\date{}

\begin{document}
\title{Joint Activity Detection and Channel Estimation in Cell-Free Massive MIMO Networks with Massive Connectivity}
\author{\IEEEauthorblockN{Mangqing Guo and M. Cenk Gursoy} \\
\thanks{The authors are with the Department of Electrical Engineering and Computer Science, Syracuse University, Syracuse, NY 13244.
Email: mguo06@syr.edu, mcgursoy@syr.edu
}
%\thanks{This work was supported by xxx.}% <-this % stops a space
}
\maketitle
%\vspace{-1.2cm}
\begin{abstract}
Cell-free massive MIMO is one of the key technologies for future wireless communications, in which users are simultaneously and jointly served by all access points (APs). In this paper, we investigate the minimum mean square error (MMSE) estimation of effective channel coefficients in cell-free massive MIMO systems with massive connectivity. To facilitate the theoretical analysis, only single measurement vector (SMV) based MMSE estimation is considered in this paper, i.e., the MMSE estimation is performed based on the received pilot signals at each AP separately. Inspired by the decoupling principle of replica symmetric postulated MMSE estimation of sparse signal vectors with independent and identically distributed (i.i.d.) non-zero components, we develop the corresponding decoupling principle for the SMV based MMSE estimation of sparse signal vectors with independent and non-identically distributed (i.n.i.d.) non-zero components, which plays a key role in the theoretical analysis of SMV based MMSE estimation of the effective channel coefficients in cell-free massive MIMO systems with massive connectivity. Subsequently, based on the obtained decoupling principle of MMSE estimation, likelihood ratio test and the optimal fusion rule, we perform user activity detection based on the received pilot signals at only one AP, or cooperation among the entire set of APs for centralized or distributed detection. Via theoretical analysis, we show that the error probabilities of both centralized and distributed detection tend to zero when the number of APs tends to infinity while the asymptotic ratio between the number of users and pilots is kept constant. We also investigate the asymptotic behavior of oracle estimation in cell-free massive MIMO systems with massive connectivity via random matrix theory. Moreover, in order to demonstrate the potential performance loss of SMV based MMSE estimation, which does not employ the correlation between the received pilot signals at different APs, the multiple measurement vector (MMV) based MMSE estimation, i.e., joint MMSE estimation with pilot signals from all APs, is analyzed via numerical results. Numerical analysis shows that the theoretical analyze with our decoupling principle for the SMV based MMSE estimation of sparse signal vectors with i.n.i.d. non-zero components matches well with the numerical results.
\end{abstract}


\section{Introduction}
Internet-of-Things (IoT) is one of the key technologies for future wireless communications, and massive connectivity and sporadic activity are two main characteristics of these IoT networks\cite{ChenSohrabiYu2018,LiuYu2018}. In particular, IoT networks include large number of users. However, each user becomes active sporadically and connects to the network only when it needs to exchange information. Otherwise, it remains inactive. Therefore, at a given time instant, only a small unknown subset out of the entire set of users is active, while most of the users are inactive.
%As the spectral efficiency (SE) is critical in nowadays, the overhead of transmitting sporadic payloads from massive set of users becomes reasonable\cite{BanaCarvalho]SoretEtAl2019}.
Then, identifying the active users is a key requirement in IoT networks with massive connectivity, and this is also referred to as the activity detection problem. In addition to activity detection, it is also critical to estimate the channel coefficients between the active users and access points (APs).

In the literature, IoT networks with massive connectivity have been intensively studied. For instance, massive connectivity for industrial IoT networks has been addressed in \cite{LeeYang2020}, where the signal-to-interference ratio (SINR) and spectral efficiency (SE) under the maximum ratio (MR) or zero-forcing (ZF) decoding at the receiver are investigated for the uplink network. Massive connectivity for IoT in distributed massive MIMO networks is analyzed in \cite{YuanHeMatthaiouEtAl2020}, where the activity detection problem is addressed via the alternating direction method of multipliers (ADMM) algorithm in light-load traffic scenarios and via the dynamic clustering method in heavy-load traffic scenarios. The orthogonal matching pursuit (OMP) algorithm is used in \cite{BoljanovicVukobratovicPopovskiEtAl2017} to deal with the activity detection problem, and the authors in \cite{MonseesBockelmannDekorsy2015} have proposed a compressed sensing Neyman-Pearson based activity detection algorithm for IoT networks with massive connectivity. Activity detection using the maximum likelihood method is investigated in \cite{FenglerCaireJungEtAl2019}.

Moreover, several recent studies have analyzed the joint activity detection and effective channel estimation (JADECE) in IoT networks with massive connectivity by employing the approximate message passing (AMP) algorithm. AMP algorithm can obtain the minimum mean square error (MMSE) estimate of sparse signals with tolerable complexity, and it breaks down the received signal into Gaussian noise corrupted scalar versions. Besides, the variance of the Gaussian noise in the decoupled signal can be described with the state evolution equations, facilitating the performance analysis of the AMP algorithm. With these benefits, AMP algorithm has been widely used in solving the JADECE problem in IoT networks with massive connectivity. For instance, the AMP algorithm is used in \cite{ChenSohrabiYu2018,LiuYu2018,LiuLarssonYuEtAl2018} to solve the JADECE problem in single-cell massive MIMO networks with massive connectivity, and multi-cell massive MIMO cooperative networks are analyzed with the AMP algorithm in \cite{ChenSohrabiYu2018a}. Authors in \cite{HanNiuRenEtAl2014,HanNiuRen2014,HayakawaNakaiHayashi2018} focus on solving the JADECE problem via the distributed AMP algorithm. The authors in \cite{JiangShiZhangEtAl2019} have proposed a structured group sparsity estimation approach to solve the JADECE problem in IoT networks with massive connectivity. Furthermore, AMP algorithm has been used in \cite{XiaShiZhouEtAl2021} to solve the JADECE problem in reconfigurable intelligent surface (RIS) assisted massive MIMO networks with massive connectivity. In addition to the AMP algorithm, sparsity learning based methods are used in \cite{DingYuanLiew2019} to deal with the JADECE problem in grant-free massive-device multiple access system. The authors in \cite{YangJinWenEtAl2019} employ a compressed sensing algorithm to solve the JADECE problem for massive connectivity with 1-bit digital-to-analog converter. A turbo receiver based on the bilinear generalized approximate message passing algorithm is proposed to solve the JADECE problem for grant-free massive random access in \cite{BianMaoZhang2021}.

Although AMP algorithm is efficient in solving the JADECE problem, it is sensitive to the mean and condition number of the measurement matrix, which impairs the convergence property of the algorithm \cite{CaltagironeZdeborovKrzakala2014,RanganSchniterFletcherEtAl2014,BirgmeierGoertz2019}. In order to overcome this, various modified AMP algorithms have been proposed, including mean removal AMP\cite{BirgmeierGoertz2019}, swept AMP\cite{Mano2015}, and complex Bayesian approximate message passing (CB-AMP) algorithm \cite{MengWuKuangEtAl2015}. Among these algorithms, CB-AMP has an improved convergence property and has state evolution equations similar to those of the AMP algorithm \cite{KrzakalaMezardSaussetEtAl2012}, rendering it a preferable choice to employ in diverse settings.

However, the state evolution equations of CB-AMP algorithm depend on the asymptotic mean square error of the estimated signal. Therefore, in order to theoretically analyze the performance of CB-AMP algorithm, one needs to identify the corresponding asymptotic mean square error first. Since we consider MMSE estimation here, how to find the asymptotic mean square error of the MMSE estimate of the sparse signal is a core problem that we need to address in this paper.

Replica method from statistical physics has been widely used to find the asymptotic mean square error of the AMP algorithm\cite{BayatiMontanari2011,HannakPerelliGoertzEtAl2018}, and it is shown to work well empirically. With the replica method, the authors in \cite{DongningGuo2005} have provided a decoupling principle for the asymptotic analysis of MMSE estimation in large system limit, and the replica analysis for MMSE estimation of multimeasurement vector (MMV) problem is performed in\cite{ZhuBaronKrzakala2017}. Moreover, the asymptotic behavior of MMSE estimation for sparse signals with random matrix theory and large deviations is analyzed in\cite{HuleihelMerhav2017,Tulino2013}.

Cell-free massive MIMO is another promising technology for future wireless communications, in which a large number of APs are distributed in an area, and all the users are simultaneously and jointly served by the entire set of APs. The channel hardening and favorable propagation properties of cell-free massive MIMO networks are studied in \cite{ChenBjrnson2018}. The achievable rates of cell-free massive MIMO systems are investigated in\cite{BasharCumananBurrEtAl2018a,NgoAshikhminYangEtAl2015,BasharCumananBurrEtAl2018,InterdonatoNgoLarssonEtAl2016,NgoAshikhminYangEtAl2017}. Power control in cell-free massive MIMO networks is studied in \cite{BuzziZappone2017,BoroujerdiAbbasfarGhanbari2017,NayebiAshikhminMarzettaEtAl2015}. Spectral and energy efficiency of cell-free massive MIMO networks are analyzed in\cite{NguyenDuongNgoEtAl2017,NgoTranDuongEtAl2018,NayebiAshikhminMarzettaEtAl2017,LiuLuoChenEtAl2020,ZhangWeiBjrnsonEtAl2017}.
We recently studied activity detection in cell-free massive MIMO systems in \cite{Guo-GlobalSIP19, Guo-ISIT20} using the AMP algorithm. Cloud-RAN (cloud radio access network) is another distributed network architecture which moves computations from each AP to the baseband unit (BBU) pool\cite{ShiZhangLetaief2014}. Compared with Cloud-RAN, each AP in cell-free massive MIMO networks has its own signal processing ability, and the received signals at each AP can be preprocessed to reduce the transmitting overhead between APs and the central processing unit (CPU).

In this paper, we focus on the asymptotic analysis of single measurement vector (SMV) based MMSE estimation in cell-free massive MIMO systems with massive connectivity. We address the JADECE problem using the frameworks of MMSE estimation and likelihood ratio test. Prior studies on the MMSE estimation of sparse signals have developed methodologies, including a closed-form method in \cite{ProtterYavnehElad2010}, an approximation method in \cite{Larsson2007}, and AMP and CB-AMP algorithms. Since we consider cell-free massive MIMO in this paper, the non-zero components of the sparse signal vectors to be estimated are independent and non-identically distributed (i.n.i.d.). However, to the best of our knowledge, all the asymptotic analyses of the MMSE estimation with replica method in the literature focus on the sparse signal vectors with independent and identically distributed (i.i.d.) non-zero components. With i.n.i.d. non-zero components, analysis becomes more challenging. Although the authors in \cite{ProtterYavnehElad2010}  have obtained closed-form expressions for the MMSE estimates of the sparse signals with heteroscedastic non-zero entries, there is a requirement for the measurement matrix to be unitary, while we in our setting have an underdetermined system as a result of the facts that limited number of pilots can be used, and condition of having a unitary measurement matrix cannot be satisfied. An approximation method for the MMSE estimation of sparse signals is provided in\cite{Larsson2007}. However, when the number of variables becomes large, the computational complexity will grow quickly and the accumulated error will be high.

Within the considered setting, our contributions in this paper can be summarized as follows:
\begin{itemize}
\item For the theoretical analysis, we consider the SMV based MMSE estimation, i.e., the MMSE estimation of the effective channel coefficients in cell-free massive MIMO networks with massive connectivity is performed with the received pilot signals at different APs separately. For the MMV based MMSE estimation, i.e., the MMSE estimation being performed jointly with the received pilot signals at all APs, we conduct a numerical analysis and demonstrate performance differences with that of SMV based MMSE estimation via numerical results.

\item Inspired by the decoupling property of replica symmetric postulated MMSE estimation for sparse signal vectors with i.i.d. non-zero components\cite{DongningGuo2005,Rangan2012}, we establish a decoupling principle of SMV based MMSE estimation for sparse signal vectors with i.n.i.d. non-zero components, which plays a key role in our theoretical analysis of SMV based MMSE estimation in cell-free massive MIMO networks with massive connectivity.

\item Using the decoupling principle, likelihood ratio test and the optimal fusion rule, we obtain detection rules for the activity of users based on the received pilot signals at only one AP, and also based on the cooperation of the received pilot signals from the entire set of APs for centralized and distributed detection.

\item We determine the false alarm and miss detection probabilities of activity detection in cell-free massive MIMO networks with massive connectivity. We show that the error probabilities of both centralized and distributed activity detection schemes tend to zero when the number of APs tends to infinity while the asymptotic ratio between the number of users and pilots is kept constant.

\item Moreover, we analyze oracle estimation, which provides a lower bound on the mean square error of MMSE estimation for sparse signals, as a benchmark scheme in the case of known user activities. In particular, we investigate the asymptotic behavior of oracle estimation in cell-free massive MIMO systems with massive connectivity via random matrix theory, and provide comparisons.
\end{itemize}

The organization of the remainder of this paper is as follows. System model of cell-free massive MIMO networks with massive connectivity is introduced in Section \uppercase\expandafter{\romannumeral2}. Several preliminary results on postulated MMSE estimation and CB-AMP algorithm are given in Section \uppercase\expandafter{\romannumeral3}. Oracle and SMV based MMSE estimation of sparse signals in cell-free massive MIMO networks with massive connectivity are analyzed in Section \uppercase\expandafter{\romannumeral4}. The analysis of the likelihood ratio test based on the decoupling principle of the SMV based MMSE estimation for sparse signal vectors with i.n.i.d. non-zero components is conducted in Section \uppercase\expandafter{\romannumeral5}. And the centralized and distributed activity detection schemes based on the decoupling principle, likelihood ratio test and the optimal fusion rule are derived in Section \uppercase\expandafter{\romannumeral6}. Extensions to MMV based MMSE estimation and activity detection is discussed in Section \uppercase\expandafter{\romannumeral7}. Numerical results are presented in Section \uppercase\expandafter{\romannumeral8} and the paper is finally concluded in Section \uppercase\expandafter{\romannumeral9}.

\section{System Model}
\begin{figure}[htbp]
	\center
	\includegraphics[width=3.3in]{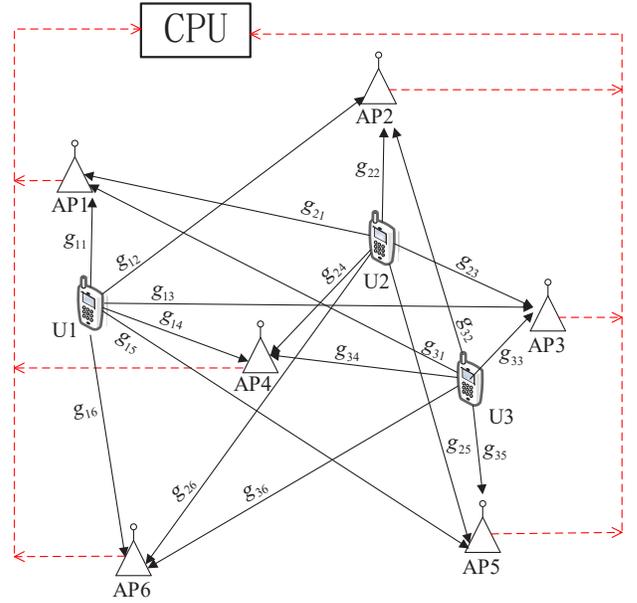}\\
	\caption{Cell-free massive MIMO network.}\label{fig0}
\end{figure}

We consider a cell-free massive MIMO network that consists of $M$ APs and $N$ single antenna users as depicted in Fig. \ref{fig0}. There is only one antenna at each AP. All the APs and users are uniformly distributed in a circular area with radius $R$, and all the APs are connected to the CPU through a backhaul network \cite{AfshangDhillon2017}. We consider sparse activity detection in the presence of massive connectivity, i.e., $N$ is very large, and among all the users, only a small fraction of them are active at a given time instant. We assume the probability of each user being active is $\lambda $, and the user activities remain the same during each channel coherence interval.

We assume flat fading channel between all the users and APs. The channel coefficient from the $i$th user to the $j$th AP is
\begin{equation}\label{equ1}
  {g_{\textit{ij}}} = \beta _{\textit{ij}}^{1/2}{h_{\textit{ij}}}
\end{equation}
where ${\beta _{\textit{ij}}}$ is the large-scale fading coefficient which can be expressed as ${\beta _{\textit{ij}}} = \min \left( {d_{\textit{ij}}^{ - \alpha },d_0^{ - \alpha }} \right)$, where ${d_{\textit{ij}}}$ is the distance from the $i$th user to the $j$th AP, ${d_0}$ is the reference distance, and $\alpha $ is the path loss decay exponent \cite{ChenBjrnson2018}. As all the APs and users are uniformly distributed in a circle with radius $R$, the probability density function of ${d_{\textit{ij}}}$ is
\begin{equation}\label{equ01}
p({d_{\textit{ij}}}) = \frac{{4{d_{\textit{ij}}}}}{{\pi {R^2}}}\left[ {{{\cos }^{ - 1}}\left( {\frac{{{d_{\textit{ij}}}}}{{2R}}} \right) - \frac{{{d_{\textit{ij}}}}}{{2R}}{{\left( {1 - \frac{{d_{\textit{ij}}^2}}{{4{R^2}}}} \right)}^{\frac{1}{2}}}} \right]
\end{equation}
for $0 < {d_{\textit{ij}}} < 2R$, and $p({d_{\textit{ij}}})=0$ elsewhere \cite{Mathai1999}.

We denote the probability density function of ${\beta _{\textit{ij}}}$ by $p(\beta )$. Since ${\beta _{\textit{ij}}}$ changes slowly over time, we assume it is known at CPU and all APs. We denote by ${{\bm{\beta }}_j} \in {{\cal R}^{N \times 1}}$ the large-scale fading vector from all users to the $j$th AP. Hence, the $i$th element of ${{\bm{\beta }}_j}$ equals ${\beta _{\textit{ij}}}$. We define ${{\bf{\Lambda }}_j} = \textit{diag}\left( {{{\bm{\beta }}_j}} \right)$. Furthermore, ${h_{\textit{ij}}} \sim {\cal C}{\cal N}(0,1)$ is the corresponding small-sale fading coefficient from the $i$th user to the $j$th AP, and all the small-scale fading coefficients are independent and identically distributed. We use ${{\bf{h}}_j}$ to denote the small-scale fading coefficients from all users to the $j$th AP.

We describe the activity of the $i$th user with binary-valued $a_i$, i.e., $a_i=1$ if the $i$th user is active, while $a_i=0$ if it is inactive. The probability of each user being active is $\lambda $, i.e., $P({a_i} = 1) = \lambda $ and $P({a_i} = 0) = 1-\lambda $. We designate the combination of the $i$th user's activity and the channel coefficient from the $i$th user to the $j$th AP as the effective channel coefficient ${\theta _{\textit{ij}}}$, i.e., ${\theta_{\textit{ij}}} = {a_i}{g_{\textit{ij}}}$. Additionally, we denote by ${b_{\textit{ij}}} = {a_i}{h_{\textit{ij}}}$ the effective small-scale fading coefficient from the $i$th user to the $j$th AP. Similarly, we use the vectors ${{\bm{\theta }}_j}$ and ${{\bf{b}}_j}$ to denote the effective channel coefficients and effective small-scale fading coefficients from all users to the $j$th AP. It is obvious that ${{\bm{\theta }}_j} = {{\bf{\Lambda }}_j}{{\bf{b}}_j}$. We indicate the probability density function of ${{\bm{\theta }}_j}$ as $p({\bm{\theta }_j})$.

In each channel coherence interval, active users will send pilots to APs first, and then transmit messages. The pilots are used by APs to recognize the activity of the users. We assume that $L$ symbols are used for pilot transmission during each channel coherence interval. The pilot matrix is denoted by an $L \times N$ complex matrix ${\bf{\Phi }}$, and each element of ${\bf{\Phi }}$ is independent and circularly symmetrically distributed with zero mean and variance $\frac{1}{L}$. Then, the received pilot signal at the $j$th AP is
\begin{IEEEeqnarray}{rCl}\label{equ02}
 {{\bf{y}}_j} &=& {\bf{\Phi }}{{\bm{\theta }}_j} + {{\bf{n}}_j} \IEEEyesnumber\IEEEyessubnumber*\\
  &=&{\bf{\Phi }}{{\bf{\Lambda }}_j^{1/2}}{{\bf{b}}_j} + {{\bf{n}}_j} \IEEEyessubnumber
\end{IEEEeqnarray}
where ${{\bf{n}}_j} \in {\cal C}{\cal N}(0,{\sigma _0^2}{{\bf{I}}_L})$ is the i.i.d. additive white Gaussian noise vector at the $j$th AP.

Finally, the aggregated received pilot signal of the entire set of APs can be expressed as
 \begin{equation}\label{equ4}
   {\bf{Y}} = {\bf{\Phi \Theta }} + {\bf{N}}
 \end{equation}
 where ${\bf{\Theta }} = [{{\bm{\theta }}_1},{{\bm{\theta }}_2}, \cdots ,{{\bm{\theta }}_M}]$ is the effective channel matrix from all users to the entire set of APs, and ${\bf{N}} = [{{\bf{n}}_1},{{\bf{n}}_2}, \cdots ,{{\bf{n}}_M}]$ is the aggregated noise matrix at all APs.

 \textbf{Remark 1:} For facilitating the theoretical analysis, we consider SMV based MMSE estimation of ${{\bm{\theta }}_j}$ in (3a), although the estimation of ${\bf{\Theta }}$ in (\ref{equ4}) is an MMV problem. The performance gap introduced by dividing the MMV problem into multiple SMV problems is shown via numerical results in Section \uppercase\expandafter{\romannumeral8}.

 \textbf{Remark 2:} Through Section \uppercase\expandafter{\romannumeral3} to Section \uppercase\expandafter{\romannumeral5}, we consider SMV based MMSE estimation and activity detection with the received pilot signals at only one AP. For instance, if we consider the $j$th AP, then the large-scale fading coefficients, effective channel coefficients, effective small-scale fading coefficients, received pilot signals and noise are ${{\bf{\Lambda }}_j}$, ${{\bm{\theta }}_j}$, ${{\bf{b}}_j}$, ${{\bf{y}}_j}$ and ${{\bf{n}}_j}$, respectively. For the sake of notational brevity, we omit the subscript ``$j$'' in these symbols through Section \uppercase\expandafter{\romannumeral3} to Section \uppercase\expandafter{\romannumeral5} when there is no confusion.

 For the completeness of the paper, in the next section, we briefly address preliminary characterizations on SMV based postulated MMSE estimation and the CB-AMP algorithm, which will be used in our subsequent analysis.

\section{Preliminaries}
 Several preliminary characterizations from prior work are provided in this section. First, we introduce the decoupling principle of SMV based postulated MMSE estimation of sparse signal vectors with i.i.d. non-zero components. Then, we describe the SMV based CB-AMP algorithm for MMSE estimation of sparse signals.
 \subsection{ SMV Based Postulated MMSE Estimation}
 From Section \uppercase\expandafter{\romannumeral2}, we know that the elements of ${\bf{b}}$ are i.i.d., and the variance of the AWGN noise at AP is $\sigma _0^2$. We denote the probability density function of each element of ${\bf{b}}$ by ${p_0}(b)$. Now we can express
 \begin{equation}\label{equ3_1}
 {p_0}(b) = \left( {1 - \lambda } \right)\delta \left( b \right) + \lambda {\cal C}{\cal N}\left( {b;0,1} \right)
 \end{equation}
 where $\delta ( \cdot )$ is the Dirac delta function, and ${\cal C}{\cal N}\left( {{x_0};{c_0},{d_0}} \right)$ stands for the probability density function (at point $x_0$) of a circularly symmetric complex Gaussian random variable $x$ with mean $c_0$ and variance $d_0$.

 Suppose we are given the postulated prior distribution of each element of ${\bf{b}}$ as ${p_{\textit{post}}}(b)$, and the postulated AWGN noise variance is $\sigma _{\textit{post}}^2$. Then, the postulated MMSE (PMMSE) estimate of ${\bf{b}}$ is defined as
 \begin{equation}\label{equ3_2}
 {\widehat {\bf{b}}^{\textit{ pmmse}}}\left( {\bf{y}} \right) = E\left\{ {{\bf{b}}|{\bf{y}};{p_{\textit{post}}}(b),\sigma _{\textit{post}}^2} \right\}.
 \end{equation}
Note that in the case in which ${p_{\textit{post}}}(b) = {p_0}(b)$ and $\sigma _{\textit{post}}^2 = \sigma _0^2$, ${\widehat {\bf{b}}^{\textit{ pmmse}}}\left( {\bf{y}} \right)$ is the MMSE estimate of ${\bf{b}}$ given the received signal ${\bf{y}}$. Thus, postulated MMSE estimation is used in this paper to aid the asymptotic analysis of SMV based MMSE estimation for sparse signal vectors with both i.i.d. and i.n.i.d. non-zero components.

 Now, let us consider a scalar Gaussian noise corrupted received signal
 \begin{equation}\label{equ3_3}
 z = b + \sqrt \eta  n
 \end{equation}
 where $n \sim {\cal C}{\cal N}(0,1)$. We define the conditional MMSE estimate of $b$ as
 \begin{equation}\label{equ3_4}
 \widehat b_{\textit{scalar}}^{\textit{ pmmse}}\left( {z;{p_{\textit{post}}}(b),\eta } \right) = E\left\{ {b|z;{p_{\textit{post}}}(b),\eta } \right\}.
 \end{equation}
 Given two distributions, $p_{\textit{post}}^1(b)$ and $p_{\textit{post}}^2(b)$, and two noise levels, ${\eta _1}$ and ${\eta _2}$, we define
 \begin{align}\label{equ3_5}
 & \textit{MSE}\left( {p_{\textit{post}}^1(b),p_{\textit{post}}^2(b),{\eta _1},{\eta _2}} \right)  \nonumber\\
 =& {\int {\left| {b - \widehat b_{\textit{scalar}}^{\textit{ pmmse}}\left( {z;p_{\textit{post}}^1(b),{\eta _1}} \right)} \right|} ^2}p\left( {b|z;p_{\textit{post}}^2(b),{\eta _2}} \right)db.
\end{align}
Equation (\ref{equ3_5}) shows the mean square error of postulated MMSE estimation for signal distribution $p_\textit{post}^2(b)$ and noise variance ${\eta _2}$ under postulated signal distribution $p_\textit{post}^1(b)$ and postulated noise variance ${\eta _1}$. When $p_\textit{post}^1(b)=p_\textit{post}^2(b)=p_0(b)$ and ${\eta _1}={\eta _2}=\sigma _0^2$, equation (\ref{equ3_5}) gives the mean square error of MMSE estimation for $\bf{b}$. Then, we can obtain the decoupling principle of replica symmetric PMMSE estimation for sparse signal vectors with i.i.d. non-zero components. We present this characterization as a property below.

 \begin{myclm}
 	\textit{[Decoupling Principle of Replica Symmetric SMV based MMSE Estimation for Sparse Signal Vectors with i.i.d. non-zero components]\cite{DongningGuo2005,Rangan2012}:} Consider the linear model in (3b). Assume that the number of pilots used for estimation, $L$, changes with the number of users, $N$ (using $L(N)$ to denote the relationship between $L$ and $N$), and $\mathop {\lim }\limits_{N \to \infty } \frac{N}{{L(N)}} = \gamma $. Let ${\widehat {\bf{b}}^{\textit{ pmmse}}}\left( {\bf{y}} \right)$ be the MMSE estimate of ${\bf{b}}$ based on the prior distribution ${p_0}(b)$ and noise variance $\sigma _0^2$. Then, under replica symmetry, there exist effective noise levels $\sigma _{\textit{eff}}^2$ and $\sigma _{\textit{p-eff}}^2$ such that: \\
 	(1) As ${N \to \infty }$, the random vector ($b_k$,${\beta}_k$,$\widehat b_k^{\textit{ pmmse}}$) converges in distribution to the random vector ($b$,$\beta$,$\widehat b$). Here, $b_k$ and $\widehat b_k^{\textit{ pmmse}}$ are the $k$th elements of ${\bf{b}}$ and ${\widehat {\bf{b}}^{\textit{ pmmse}}}\left( {\bf{y}} \right)$, respectively. ${\beta}_k$ is the $k$th diagonal element of ${\bf{\Lambda }}$. $b$, $\beta$ and $n$ are independent with distributions $b \sim {p_0}(b)$, $\beta  \sim p(\beta )$, $n \sim {\cal C}{\cal N}(0,1)$, and
 	\begin{IEEEeqnarray}{rCl}\label{equ3_6}
 		&z^{\textit{mmse}} = b + \sqrt \eta  n, \IEEEyesnumber\IEEEyessubnumber*\\
 		&\widehat b = \widehat b_{\textit{scalar}}^{\textit{ pmmse}}\left( {z^{\textit{mmse}};{p_0}(b),{\eta _p}} \right) \IEEEyessubnumber
 	\end{IEEEeqnarray}
    where $\eta  = \frac{{\sigma _{\textit{eff}}^2}}{\beta }$ and ${\eta _p} = \frac{{\sigma _{\textit{p-eff}}^2}}{\beta }$. \\
    (2) The effective noise levels satisfy the equations
    \begin{IEEEeqnarray}{rCl}\label{equ3_7}
    	&\sigma _{\textit{eff}}^2 = \sigma _0^2 + \gamma E\left\{ {\beta \textit{MSE}\left( {{p_0}(b),{p_0}(b),{\eta _p},\eta ,z} \right)} \right\}, \IEEEyesnumber\IEEEyessubnumber*\\
    	&\sigma _{\textit{p-eff}}^2 = \sigma _0^2 + \gamma E\left\{ {\beta \textit{MSE}\left( {{p_0}(b),{p_0}(b),{\eta _p},{\eta _p},z} \right)} \right\} \IEEEyessubnumber
    \end{IEEEeqnarray}
    where the expectations are taken over $\beta  \sim p(\beta )$ and $z$ is generated by (10a).	
 \end{myclm}

Property 1 shows that the MMSE estimation of the SMV problem in (3b) can be decoupled into scalar Gaussian noise corrupted linear problems as described in (10a), and the corresponding noise variance in (10a) can be obtained by solving the fixed point equations in (11). Then, we can readily find the MMSE estimates in (10b) and the corresponding mean square error.

\textbf{Remark 3:} Since we consider SMV based MMSE estimation in this paper, the postulated prior of ${\bf{b}}$ and the postulated noise level used in Property 1 are the prior distribution ${p_0}(b)$ and noise variance $\sigma _0^2$, respectively.

\textbf{Remark 4:} In general, there may exist multiple solutions for the fixed point equations in (\ref{equ3_7}). In this case, the true solution is the minimizer of a certain free energy function as described in \cite{DongningGuo2005}.

 \subsection{ SMV Based CB-AMP Algorithm}
 The CB-AMP algorithm is used during numerical simulations to obtain the SMV based MMSE estimates of the sparse effective channel coefficients in cell-free massive MIMO networks. For the sake of completeness in the paper, we briefly introduce the CB-AMP algorithm proposed in \cite{MengWuKuangEtAl2015}. We summarize the recursions of the CB-AMP algorithm in Algorithm 1 below.
 \begin{algorithm}
 	\caption{SMV based CB-AMP algorithm \cite{MengWuKuangEtAl2015}.}
 	\label{alg1}
 	\begin{flushleft}
 		For the linear model in (3a) with a given pilot matrix ${\bf{\Phi }}$, received signal ${\bf{y}}$, and the prior probability function ${p({\bm{\theta }})}$, the CB-AMP algorithm generates a sequence of estimates ${\widehat {\bm{\theta }}^t}$, ${\widehat {\bf{r}}^t}$, for $t = 1,2, \ldots $ through the following recursions:
 	\end{flushleft}
 	\begin{algorithmic}
 		\STATE (1) Initialization: Set $t = 1$, ${\widehat {\bm{\theta }}^1} = \overline {\bm{\theta }} $, ${\widehat {\bm{\kappa }}^1} = \operatorname{var} ({\bm{\theta }})$, ${{\bf{z}}^0} = {{\bf{1}}_L}$, ${\bf{p}} = {\bf{y}}$, where $\overline {\bm{\theta }} $ and $\operatorname{var} ({\bm{\theta }})$ are the corresponding mean and variance for each element of ${\bm{\theta }}$, and ${{\bf{1}}_M}$ is an $1 \times M$ vector whose elements are 1.
 		\STATE (2) For each $j \in [L]$ (where $[L]$ stands for the set which includes the integers from 1 to $L$), calculate
 		\begin{IEEEeqnarray}{rCl}\label{equ3_16}
 			&z_j^t = \sum\limits_i {{{\left| {{\phi _{\textit{ji}}}} \right|}^2}{\widehat {\kappa }} _i^t}, \IEEEyesnumber\IEEEyessubnumber*\\
 			&p_j^t = \sum\limits_i {{\phi _{\textit{ji}}}\widehat \theta _i^t}  - \frac{{z_j^t}}{{{\sigma_0 ^2} + z_j^{t - 1}}}\left( {{y_a} - p_a^{t - 1}} \right) \IEEEyessubnumber
 		\end{IEEEeqnarray}
 	    where ${{\phi _{\textit{ji}}}}$ is the corresponding element on the $j$th row and $i$th column of ${\bf{\Phi }}$.
       \STATE (3) For each $i \in [N]$, calculate
       \begin{IEEEeqnarray}{rCl}\label{equ3_18}
     	  &\tau _i^t = {\left[ {\sum\limits_j {\frac{{{{\left| {{\phi _{\textit{ji}}}} \right|}^2}}}{{{\sigma_0 ^2} + z_j^t}}} } \right]^{ - 1}}, \IEEEyesnumber\IEEEyessubnumber*\\
     	  &\widehat r_i^t = \widehat \theta _i^t + \tau _i^t\sum\limits_j {\frac{{\phi _{\textit{ji}}^*\left( {{y_j} - p_j^t} \right)}}{{{\sigma_0 ^2} + z_j^t}}},\IEEEyessubnumber\\
     	  &\widehat \theta _i^{t + 1} = {g_{\textit{in}}}\left( {\widehat r_i^t,\tau _i^t} \right), \IEEEyessubnumber\\
       	  &\widehat \kappa _i^{t + 1} = {\varpi_{\textit{in}}}\left( {\widehat r_i^t,\tau _i^t} \right). \IEEEyessubnumber
       \end{IEEEeqnarray}
      Then let $t \leftarrow t+1$, and return to step (2) and repeat until a sufficient number of iterations have been performed or a given termination condition is satisfied.
 	\end{algorithmic}
 \end{algorithm}

Considering the AWGN output channel and the recursions of the CB-AMP algorithm, the estimated signal is decoupled into scalar AWGN Gaussian noise corrupted signals, i.e.,
\begin{equation}\label{equ3_20}
\widehat r_i^t = {\theta _i} + \sqrt {{\xi ^t}} n
\end{equation}
where $\xi ^t$ is the variance of the corrupting noise, and it satisfies the following state evolution equation \cite{MengWuKuangEtAl2015}:
\begin{equation}\label{equ3_21}
{\xi ^{t + 1}} = \sigma _0^2 + \gamma E\left\{ {{{| {{\theta _i} - \widehat \theta _i^t} |}^2}} \right\}.
\end{equation}

The MMSE estimates of $\widehat \theta _i^t$ given $\widehat r_i^t$ for the linear model in (\ref{equ3_20}) is \cite{Tulino2013}
\begin{align}\label{equ3_22}
\widehat \theta _i^{t + 1}\left( {\widehat r_i^t;{{\xi ^t}} ,\lambda ,{\beta _i}} \right) &= E\left\{ {\theta |\widehat r_i^t;{{\xi ^t}} ,\lambda ,{\beta _i}} \right\}  \nonumber\\
& = G\left( {{{| {\widehat r_i^t} |}^2};{{\xi ^t}} ,\lambda ,{\beta _i}} \right)\frac{{{\beta _i}}}{{{\beta _i} + {{\xi ^t}} }}\widehat r_i^t
\end{align}
where
\begin{equation*}
G\left( {{{| {\widehat r_i^t} |}^2};{{\xi ^t}} ,\lambda ,{\beta _i}} \right) = \frac{1}{{1 + \frac{{(1 - \lambda )({\beta _i} + {{\xi ^t}} )}}{{\lambda {{\xi ^t}} }}\exp \left( { - \frac{{{\beta _i}{|\widehat r_i^t{|^2}}}}{{{{\xi ^t}} ({\beta _i} + {{\xi ^t}} )}}} \right)}}.
\end{equation*}
Then we can obtain the MMSE estimate of $\widehat \theta _i^{t + 1}$ as
\begin{equation}\label{equ3_23}
g_{\textit{in}}(\widehat r_i^t,\tau _i^t) = \widehat \theta _i^{t + 1}\left( {\widehat r_i^t;\tau _i^t,\lambda ,{\beta _i}} \right)
\end{equation}
where $\tau _i^t$ provided in (13a) is the estimated value of ${{\xi ^t}}$. Finally, the estimated variance of $\widehat \theta _i^{t + 1}$ can be expressed as
\begin{align}\label{equ3_24}
{\varpi _{in}}( {\widehat r_i^t,\tau _i^t} ) &=\tau _i^t\frac{\partial }{{\partial \widehat r_i^t}}{g_{in}}\left( {\widehat r_i^t,\tau _i^t} \right) \nonumber\\
&={\beta _i}G\left[ {\frac{{{\beta _i}( {\tau _i^t + {{| {\widehat r_i^t} |}^2}} ) + {{| {\tau _i^t} |}^2}}}{{{{( {{\beta _i} + \tau _i^t} )}^2}}} - G\frac{{{\beta _i}{{| {\widehat r_i^t} |}^2}}}{{{{( {{\beta _i} + \tau _i^t} )}^2}}}} \right]
\end{align}
where $G$ stands for the function ${G\left( {{{\left| {\widehat r_i^t} \right|}^2};\tau _i^t,\lambda ,{\beta _i}} \right)}$.

\section{ SMV Based MMSE Estimation in Cell-Free Massive MIMO Networks with Massive Connectivity}
In this paper, we focus on the asymptotic analysis of SMV based MMSE estimation for sparse signal vectors with i.n.i.d. non-zero components, which is the case for the linear model in (3a). Since it is challenging to find closed-form results, CB-AMP is an efficient algorithm to acquire the MMSE estimate of ${{\bm{\theta }}_j}$ numerically. However, to the best of our knowledge, there is no previous work which has focused on the asymptotic analysis of this setting. Several previous studies in the literature addressed the asymptotic analysis of SMV based MMSE estimation for sparse signal vectors with i.i.d. non-zero components. In order to use these results, we need to convert the problem in (3a) into (3b), in which case the unknown sparse vector $\bf{b}_j$ has i.i.d. non-zero components. On the other hand, in the same situation with (3a), it is challenging for us to find a closed-form MMSE estimate of $\bf{b}_j$ in (3b), and CB-AMP is an efficient algorithm to be considered. However, CB-AMP has strict requirements on the structure of the measurement matrix to ensure that the algorithm works\cite{MengWuKuangEtAl2015,KrzakalaMezardSaussetEtAl2012}. For a given pilot matrix ${\bf{\Phi }}$, the equivalent measurement matrix in (3b) is ${\bf{\Phi }}{{\bf{\Lambda }}_j}$ as we consider only ${{\bf{y}}_j}$ for SMV based MMSE estimation when seeking to estimate ${{\bf{b}}_j}$. Since the elements of ${{\bf{\Lambda }}_j}$ are random variables, it is difficult to control the structure of ${\bf{\Phi }}{{\bf{\Lambda }}_j}$ to meet the measurement matrix requirements of the CB-AMP algorithm. Thus, it is challenging to find the MMSE estimate of $\bf{b}_j$ in (3b) numerically with the CB-AMP algorithm. In summary, we can not perform asymptotic and numerical analysis with the same linear model, i.e., linear model in (3a) is suitable for numerical analysis but challenging for asymptotic analysis, while (3b) is challenging for numerical analyze but suitable for asymptotic analysis.

In this section, by considering the relationship between the MMSE estimate of unknown vectors in (3a) and (3b), we arrive at a decoupling principle as represented in Property 2, with which we can perform asymptotic analysis on the MMSE estimate of ${{\bm{\theta }}_j}$ in (3a). Then, the asymptotic and numerical analyses can be performed within the same linear model (3a), facilitating the performance analysis on SMV based MMSE estimation for sparse signal vectors with i.n.i.d. non-zero components (the case of cell-free massive MIMO being included), and also the likelihood ratio test in the following sections.

Initially, to provide a benchmark, we analyze the ``oracle estimation'' with which the MMSE estimation is performed using the linear model (3a) under the idealistic assumption that the support of the users (i.e., their being active or inactive) is known at the APs. With the prior information on user support, it is obvious that the oracle estimation leads to a lower bound on the mean square error of MMSE estimation using the linear model (3a) with unknown user support. Additionally, the asymptotic analyze of oracle estimation is performed with random matrix theory.

In order for the simplicity of descriptions, we define several notations first. Let us denote the user activity support of ${\bm{\theta }}$ as ${{\bf{s}}^*} = {[{a_1},{a_2}, \cdots ,{a_N}]^T}$. For a given user activity support ${\bf{s}} \in {\{ 0,1\} ^N}$ and its corresponding index set of non-zero elements, ${I_{\bf{s}}}$ ($i \in {I_{\bf{s}}}$ if $i \leqslant N$ and ${s_i}=1$), we define ${{\bf{\Phi }}_{{{\bf{s}}}}}$ and ${{\bf{\Lambda }}_{{{\bf{s}}}}}$ as matrices consisting of the columns of ${\bf{\Phi }}$ and ${\bf{\Lambda }}$ on the index set ${I_{\bf{s}}}$, respectively. Similarly, ${{\bm{\theta }}_{{{\bf{s}}}}}$ is comprised of the entries on the index set ${I_{\bf{s}}}$.

\subsection{Oracle Estimation}
Assume ${{\bf{s}}^*}$ is known at all APs. Then,
\begin{equation}\label{oracle1}
p({\bf{y}}|{{\bf{\Phi }}_{{{\bf{s}}^*}}},{{\bm{\theta }}_{{{\bf{s}}^*}}}) = \frac{1}{{{\pi ^L}{\sigma_0 ^{2L}}}}\exp \left( { - \frac{{{{\left\| {{\bf{y}} - {{\bf{\Phi }}_{{{\bf{s}}^*}}}{{\bm{\theta }}_{{{\bf{s}}^*}}}} \right\|}^2}}}{{{\sigma_0 ^2}}}} \right)
\end{equation}
and
\begin{equation}\label{oracle2}
p({{\bm{\theta }}_{{{\bf{s}}^*}}}) = \frac{1}{{{\pi ^{\left| {{{\bf{s}}^*}} \right|}}\left| {{{\bf{\Lambda }}_{{{\bf{s}}^*}}}} \right|}}\exp \left( { - {\bm{\theta }}_{{{\bf{s}}^*}}^H{\bf{\Lambda }}_{{{\bf{s}}^*}}^{ - 1}{{\bm{\theta }}_{{{\bf{s}}^*}}}} \right)
\end{equation}
where $|{{\bf{s}}^*}|$ stands for the cardinality of ${{\bf{s}}^*}$, and $\left| {{{\bf{\Lambda }}_{{{\bf{s}}^*}}}} \right|$ denotes the determinant of ${{\bf{\Lambda }}_{{{\bf{s}}^*}}}$. Therefore, the oracle estimate is
\begin{align}\label{oracle4}
{\widehat {\bm{\theta }}_{\textit{oracle}}} &= \int {{{\bm{\theta }}_{{{\bf{s}}^*}}}} p({{\bm{\theta }}_{{{\bf{s}}^*}}}|{{\bf{\Phi }}_{{{\bf{s}}^*}}},{\bf{y}})d{{\bm{\theta }}_{{{\bf{s}}^*}}}  \nonumber\\
&  = {\left( {{\bf{\Phi }}_{{{\bf{s}}^*}}^H{{\bf{\Phi }}_{{{\bf{s}}^*}}} + {\sigma_0 ^2}{\bf{\Lambda }}_{{{\bf{s}}^*}}^{ - 1}} \right)^{ - 1}}{\bf{\Phi }}_{{{\bf{s}}^*}}^H{\bf{y}}
\end{align}
and the corresponding mean square error is
\begin{equation}\label{oracle5}
{\textit{MSE}_{\textit{oracle}}} = \frac{1}{N}tr\left( {{{\left( {\frac{1}{{\sigma _0^2}}{\bf{\Phi }}_{{{\bf{s}}^*}}^H{{\bf{\Phi }}_{{{\bf{s}}^*}}} + {\bf{\Lambda }}_{{{\bf{s}}^*}}^{ - 1}} \right)}^{ - 1}}} \right)
\end{equation}
where $tr( \cdot )$ is the trace operator. Now, let us consider the asymptotic behavior of the mean square error of the oracle estimate, i.e., determine ${\textit{MSE}_{{m^*}}} = \mathop {\lim }\limits_{N \to \infty } {\textit{MSE}_{\textit{oracle}}}$, where ${m^*} =\mathop {\lim }\limits_{N \to \infty } \frac{| {{{\bf{s}}^*}}|}{N} $. Recall that the elements of ${{{\bf{\Phi }}_{{{\bf{s}}^*}}}}$ are i.i.d. with zero mean and variance $1/L$. Then we can obtain
\begin{equation}\label{asym1}
{{\cal R}_{{\frac{1}{{\sigma _0^2}}{\bf{\Phi }}_{{{\bf{s}}^*}}^H{{\bf{\Phi }}_{{{\bf{s}}^*}}}}}}\left( z \right) = \frac{1}{{\sigma _0^2 - {m^*}\gamma z}}
\end{equation}
where ${{\cal R}_{\bf{X}}}\left(  \cdot  \right)$ is the R-transform of random matrix ${\bf{X}}$\cite{Tulino2004}.

Via random matrix theory, the $\eta$-transform of ${{\bf{\Lambda }}_{{{\bf{s}}^*}}^{ - 1}}$ is
\begin{equation}\label{asym2}
{\eta _{{\bf{\Lambda }}_{{{\bf{s}}^*}}^{ - 1}}}\left(\varsigma  \right) = E\left\{ {\frac{{{\beta}}}{{{\beta } + \varsigma }}} \right\}
\end{equation}
where $\beta$ is a random variable that has the same distribution as the diagonal elements of ${{\bf{\Lambda }}_{{{\bf{s}}^*}}}$, and $\varsigma > 0$\cite{Tulino2004}. Using the relationship between R-transform and $\eta$-transform, we can obtain
\begin{equation}\label{asym3}
{{\cal R}_{{\bf{\Lambda }}_{{{\bf{s}}^*}}^{ - 1}}}\left( z \right) =  - \frac{1}{\varsigma } - \frac{1}{z}
\end{equation}
where
\begin{equation}\label{asym4}
z =  - \varsigma {\eta _{{\bf{\Lambda }}_{{{\bf{s}}^*}}^{ - 1}}}\left( \varsigma  \right).
\end{equation}

Let us define ${\bf{F}}^{\bf{s}^*} ={\frac{1}{{\sigma _0^2}}{\bf{\Phi }}_{{{\bf{s}}^*}}^H{{\bf{\Phi }}_{{{\bf{s}}^*}}} + {\bf{\Lambda }}_{{{\bf{s}}^*}}^{ - 1}}$. Since ${\frac{1}{{\sigma _0^2}}{\bf{\Phi }}_{{{\bf{s}}^*}}^H{{\bf{\Phi }}_{{{\bf{s}}^*}}}}$ is unitarily invariant and ${{\bf{\Lambda }}_{{{\bf{s}}^*}}^{ - 1}}$ is a deterministic matrix with bounded eigenvalues, ${\frac{1}{{\sigma _0^2}}{\bf{\Phi }}_{{{\bf{s}}^*}}^H{{\bf{\Phi }}_{{{\bf{s}}^*}}}}$ and ${{\bf{\Lambda }}_{{{\bf{s}}^*}}^{ - 1}}$ are asymptotically free. Thus,
\begin{align}\label{asym5}
{{\cal R}_{{\bf{F}}^{\bf{s}^*}}}\left( z \right) &= {{\cal R}_{\frac{1}{{\sigma _0^2}}{\bf{\Phi }}_{{{\bf{s}}^*}}^H{{\bf{\Phi }}_{{{\bf{s}}^*}}}}}\left( z \right) + {{\cal R}_{{\bf{\Lambda }}_{{{\bf{s}}^*}}^{ - 1}}}\left( z \right)  \nonumber\\
&  = \frac{1}{{{\sigma_0 ^2} - {m^*}\gamma z}} - \frac{1}{\varsigma } - \frac{1}{z}.
\end{align}
Therefore,
\begin{align}\label{asym6}
{\cal S}_{{\bf{F}}^{\bf{s}^*}}^{ - 1}\left( { - z} \right) &= {{\cal R}_{{\bf{F}}^{\bf{s}^*}}}\left( z \right) + \frac{1}{z}  \nonumber\\
& =\frac{1}{{{\sigma_0 ^2} - {m^*}\gamma z}} - \frac{1}{\varsigma }
\end{align}
where ${\cal S}_{\bf{X}}^{ - 1}\left(  \cdot  \right)$ stands for the inverse of Stieltjes transform of the random matrix ${\bf{X}}$. Setting ${\cal S}_{{\bf{F}}^{\bf{s}^*}}^{ - 1}\left( { - {z_0}} \right) = 0$, we can obtain
\begin{equation}\label{asym7}
E\left\{ {\frac{{{\beta}}}{{{\beta} + {\varsigma _0}}}} \right\} = \frac{{ {{\varsigma _0} - {\sigma_0 ^2}} }}{{{m^*}\gamma{\varsigma _0}}}.
\end{equation}
Then, with the definition of Stieltjes transform, we have
\begin{align}\label{asym8}
\frac{1}{{{m^*}}}{\textit{MSE}_{m^*}}&=\mathop {\lim }\limits_{| {{{\bf{s}}^*}}| \to \infty } \frac{1}{| {{{\bf{s}}^*}}|}tr\left( {{{\left( {\frac{1}{{\sigma _0^2}}{\bf{\Phi }}_{{{\bf{s}}^*}}^H{{\bf{\Phi }}_{{{\bf{s}}^*}}} + {\bf{\Lambda }}_{{{\bf{s}}^*}}^{ - 1}} \right)}^{ - 1}}} \right) \nonumber\\
&= {{\cal S}_{{\bf{F}}^{\bf{s}^*}}}\left( 0 \right)  \nonumber\\
&  = {{\cal S}_{{\bf{F}}^{\bf{s}^*}}}\left( {{\cal S}_{{\bf{F}}^{\bf{s}^*}}^{ - 1}\left( { - {z_0}} \right)} \right)  \nonumber\\
&  =  - {z_0}  \nonumber\\
&  = {\varsigma _0}{\eta _{{\bf{\Lambda }}_{\bf{s}^*}^{ - 1}}}\left( {{\varsigma _0}} \right)  \nonumber\\
&  = \frac{{ {{\varsigma _0} - {\sigma_0 ^2}} }}{{m^*}\gamma }
\end{align}
where ${{\cal S}_{\bf{X}}}\left(  \cdot  \right)$ is the Stieltjes transform of random matrix ${\bf{X}}$, and ${{\varsigma _0}}$ can be obtained via the fixed point equation (\ref{asym7}).
%Therefore,
%\begin{equation}\label{asym9}
%{\textit{MSE}_{oracle}} = \sum\limits_{k = 1}^N {\binom{N}{k}{\lambda ^k}{{(1 - \lambda )}^{N - k}}{{\textit{MSE}_{{\frac{k}{N}} }}}}.
%\end{equation}

In this paper, we consider the scenario in which the users are active independently with the same probability $\lambda$. Therefore, $\lambda = \mathop {\lim }\limits_{N \to \infty } \frac{| {{{\bf{s}}^*}}|}{N}={m^*} $. Thus, the asymptotic mean square error of the oracle estimation is
\begin{align}\label{asym9}
{\textit{MSE}_{\textit{oracle,A}}} &= MS{E_\lambda }  \nonumber\\
& = \frac{{ {{\varsigma^*} - {\sigma_0 ^2}} }}{\gamma }
\end{align}
where ${\varsigma^*}$ is the solution of the fixed point equation
\begin{equation*}
E\left\{ {\frac{{{\beta}}}{{{\beta} + {\varsigma^*}}}} \right\} = \frac{{ {{\varsigma^*} - {\sigma_0 ^2}} }}{{\lambda \gamma{\varsigma^*}}}.
\end{equation*}

\subsection{ SMV Based MMSE Estimation}
In this subsection, we focus on the asymptotic analysis of SMV based MMSE estimation for sparse signal vectors with i.n.i.d. non-zero components, which is the statistical description of linear model (3a). By comparing the formulas for the MMSE estimate of unknown vectors in (3a) and (3b), we find a closed-form relationship between the MMSE estimates of ${\bm{\theta }}$ and ${\bf{b}}$. Inspired by the decoupling principle in Property 1 for (3b), we propose a similar decoupling principle for the linear model (3a) as represented in Property 2. Finally, we arrive at two different methods to compute the effective noise level ${\sigma _{\textit{eff}}}$ for the MMSE estimates of linear model (3a).

The SMV based MMSE estimator for the unknown vector in linear model (3a) is \cite{Larsson2007}
\begin{equation}\label{mmse1}
\widehat {\bm{\theta }}_{\textit{mmse}} =\sum\nolimits_{{\bf{s}} \in {{\{ 0,1\} }^N}} {p({\bf{s}}|{\bf{y}}){\widehat {\bm{\theta }}_{\bf{s}}}}
\end{equation}
where ${p({\bf{s}}|{\bf{y}})}$ is the conditional probability that the user's activity support equals ${\bf{s}}$, and
\begin{equation}\label{mmse2}
\widehat {\bm{\theta }}_{\bf{s}} = {\left( {{\bf{\Phi }}_{\bf{s}}^H{{\bf{\Phi }}_{\bf{s}}} + \sigma _0^2{\bf{\Lambda }}_{\bf{s}}^{ - 1}} \right)^{ - 1}}{\bf{\Phi }}_{\bf{s}}^H{\bf{y}}.
\end{equation}
Similarly, the MMSE estimator for the unknown vector in linear model (3b) is
\begin{equation}\label{mmse3}
{\widehat {\bf{b}}_{\textit{mmse}}} = \sum\nolimits_{{\bf{s}} \in {{\{ 0,1\} }^N}} {p({\bf{s}}|{\bf{y}}){{\widehat {\bf{b}}}_{\bf{s}}}}
\end{equation}
and
\begin{align}\label{mmse4}
{\widehat {\bf{b}}_{\bf{s}}} &= {\left( {{\bf{\Lambda }}_{\bf{s}}^{1/2}{\bf{\Phi }}_{\bf{s}}^H{{\bf{\Phi }}_{\bf{s}}}{\bf{\Lambda }}_{\bf{s}}^{1/2} + \sigma _0^2{{\bf{I}}_{|{\bf{s}}|}}} \right)^{ - 1}}{\bf{\Lambda }}_{\bf{s}}^{1/2}{\bf{\Phi }}_{\bf{s}}^H{\bf{y}}  \nonumber\\
&= {\bf{\Lambda }}_{\bf{s}}^{-1/2}{\left( {{\bf{\Phi }}_{\bf{s}}^H{{\bf{\Phi }}_{\bf{s}}} + \sigma _0^2{\bf{\Lambda }}_{\bf{s}}^{ - 1}} \right)^{ - 1}}{\bf{\Phi }}_{\bf{s}}^H{\bf{y}}.
\end{align}

For a given received signal ${\bf{y}}$, we have the same ${p({\bf{s}}|{\bf{y}})}$ in (\ref{mmse1}) and (\ref{mmse3}), and ${\widehat {\bm{\theta }}_{\bf{s}}} = {\bf{\Lambda }}_{\bf{s}}^{1/2}{\widehat {\bf{b}}_{\bf{s}}}$. Therefore,
\begin{equation}\label{mmse5}
{\widehat {\bm{\theta }}_{\textit{mmse}}} = {{\bf{\Lambda }}^{1/2}}{\widehat {\bf{b}}_{\textit{mmse}}}.
\end{equation}

Decoupling principle in Property 1 indicates that the MMSE estimate for linear model (3b) can be decomposed into Gaussian noise corrupted scalar models described in (10a). As a result of the relationship in (\ref{mmse5}), it is natural for us to consider using ${z_1^{\textit{mmse}}} = \sqrt \beta {z^{\textit{mmse}}}$ to describe the decoupling property of the MMSE estimate for linear model (3a), i.e.,
\begin{equation}\label{mmse6}
{z_1^{\textit{mmse}}} = \theta  + {\sigma _{\textit{eff}}} n.
\end{equation}

Now, let us try to verify the assumption of (\ref{mmse6}) via the MMSE estimates of the scalar linear models in (10a) and (\ref{mmse6}). With the result in (\ref{equ3_22}), it is easy for us to obtain that
\begin{equation}\label{mmse7}
{\widehat b_{\textit{mmse}}} = G\left( {|{z^{\textit{mmse}}}{|^2};\eta ,\lambda ,1} \right)\frac{{z^{\textit{mmse}}}}{{1 + \eta }},
\end{equation}
\begin{equation}\label{mmse8}
{\widehat \theta _{\textit{mmse}}} = G\left( {|{z_1^{\textit{mmse}}}{|^2};\sigma _{\textit{eff}}^2,\lambda ,\beta } \right)\frac{{\beta {z_1^{\textit{mmse}}}}}{{\beta  + \sigma _{\textit{eff}}^2}},
\end{equation}
and
\begin{equation}\label{mmse9}
G\left( {|{z_1^{\textit{mmse}}}{|^2};\sigma _{\textit{eff}}^2,\lambda ,\beta } \right) = G\left( {|z{|^2};\eta ,\lambda ,1} \right)
\end{equation}
when $z_1^{\textit{mmse}}=\sqrt{\beta} z^{\textit{mmse}}$.
Therefore,
\begin{equation}\label{mmse10}
{\widehat \theta _{\textit{mmse}}} = \sqrt \beta  {\widehat b_{\textit{mmse}}},
\end{equation}
which is consistent with the relationship in (\ref{mmse5}) and verifies our assumption of using (\ref{mmse6}) to describe the decoupling property of the MMSE estimate for linear model (3a).

%Then, let's look at the mean squared error for the MMSE estimation of these two scalar linear models. The mean squared error for the MMSE estimation of (10a) is \cite{Tulino2013}
%\begin{equation}\label{mmse11}
%{\textit{MSE}_{\widehat b}} = \lambda \left[ {1 - \frac{{{\eta ^2}}}{{1 + \eta }}\omega \left( { - \frac{{(1 + \eta )(1 - \lambda )}}{{\eta \lambda }},2,\eta } \right)} \right]
%\end{equation}
%where
%\begin{equation*}
%\omega (a,b) = \int_0^\infty  {\frac{{t{e^{ - bt}}}}{{1 - a{e^{ - t}}}}dt}.
%\end{equation*}
%Similarly, the mean square error for the MMSE estimation of (\ref{mmse6}) is
%\begin{equation}\label{mmse12}
%{\textit{MSE}_{\widehat \theta }} = \lambda \left[ {\beta  - \frac{{\sigma _{\textit{eff}}^4}}{{\beta  + \sigma _{\textit{eff}}^2}}\omega \left( { - \frac{{(\beta  + \sigma _{\textit{eff}}^2)(1 - \lambda )}}{{\lambda \sigma _{\textit{eff}}^2}},\frac{{\sigma _{\textit{eff}}^2}}{\beta }} \right)} \right].
%\end{equation}
%Since $\eta  = \sigma _{\textit{eff}}^2/\beta $, ${\textit{MSE}_{\widehat \theta }} = \beta {\textit{MSE}_{\widehat b}}$.
Combining (\ref{mmse5}), (\ref{mmse6}) and Property 1, we conclude that the MMSE estimator of linear model (3a) breaks down the received signal into Gaussian noise corrupted versions as described in (\ref{mmse6}), and the parameter ${\sigma _{\textit{eff}}}$ satisfies the fixed point equations in (\ref{equ3_7}). We call this property as the decoupling principle of the MMSE estimation of linear model (3a) and summarize it in Property 2 as follows:

\begin{myclm}
	\textit{[Decoupling Principle of Replica Symmetric SMV based MMSE Estimation for Sparse Signal Vectors with i.n.i.d. non-zero Components]:} Consider the linear model in (3a), where ${\bm{\theta }} = {\bf{\Lambda b}}$, ${\bf{\Lambda }}$ is a diagonal matrix whose diagonal elements (denoted by the random variable $\beta$) are i.i.d. distributed with distribution $\beta  \sim p(\beta )$. The elements of $\bf{b}$ are i.i.d. distributed with distribution $b \sim {p_0}(b)$. Assume that the number of pilots used for estimation, $L$, and the number of users, $N$ satisfy the conditions as described in Property 1. Let ${\widehat {\bm{\theta }}^\textit{mmse}}({\bf{y}})$ be the MMSE estimate of ${\bm{\theta}}$. Then, under replica symmetry, as ${N \to \infty }$, the random vector ($\theta_k$,${\beta}_k$,$\widehat \theta_k^{\textit{mmse}}$) converges in distribution to the random vector ($\theta$,$\beta$,$\widehat \theta$). Here, $\theta_k$ and $\widehat \theta_k^{\textit{ mmse}}$ are the $k$th elements of ${\bm{\theta}}$ and ${\widehat {\bm{\theta}}^{\textit{ mmse}}}\left( {\bf{y}} \right)$, respectively. ${\beta}_k$ is the $k$th diagonal element of ${\bf{\Lambda }}$. $b$, $\beta$ and $n$ are independent with distributions $b \sim {p_0}(b)$, $\beta  \sim p(\beta )$, $n \sim {\cal C}{\cal N}(0,1)$. There exist effective noise levels $\sigma _{\textit{eff}}^2$ and $\sigma _{\textit{p-eff}}^2$ such that:
	\begin{IEEEeqnarray}{rCl}\label{equ3_6_1}
		&{z_1^{\textit{mmse}}} = \theta  + {\sigma _{\textit{eff}}} n, \IEEEyesnumber\IEEEyessubnumber*\\
		&\widehat \theta = \sqrt{\beta} \widehat b_{\textit{scalar}}^{\textit{ pmmse}}\left( {z_1^{\textit{mmse}};{p_0}(b),{\eta _p}} \right), \IEEEyessubnumber
	\end{IEEEeqnarray}
	where ${\eta _p} = \frac{{\sigma _{\textit{p-eff}}^2}}{\beta }$. The effective noise levels $\sigma _{\textit{eff}}^2$ and $\sigma _{\textit{p-eff}}^2$ satisfy the fixed point equations in (\ref{equ3_7}).
\end{myclm}

\begin{IEEEproof}
	From the results in Property 1, we can acquire that as $N \to \infty $, the random vector ($\frac{{{\theta _k}}}{{\sqrt {{\beta _k}} }}$,${{\beta _k}}$,$\widehat b_k^{\textit{ pmmse}}$) converges in distribution to the random vector ($\frac{\theta }{\sqrt \beta}$,$\beta$,$\widehat b$), and the scalar Gaussian noise corrupted $\theta$ can be expressed as $z^{\textit{mmse}} = \frac{\theta }{{\sqrt \beta  }} + \sqrt \eta  n$. As noted above, the MMSE estimates ${\widehat \theta _{\textit{mmse}}}$ and ${\widehat b_{\textit{mmse}}}$ have the relationship ${\widehat \theta _{\textit{mmse}}} = \sqrt \beta  {\widehat b_{\textit{mmse}}}$ when the scalar versions of Gaussian noise corrupted $\theta $ and $b$, which are denoted as $z_1^\textit{mmse}$ and $z^\textit{mmse}$, have the relationship $z_1^\textit{mmse} = \sqrt {{\beta}} {z^\textit{mmse}}$. Therefore, when the decoupled scalar Gaussian noise corrupted ${\theta}$ is expressed as $z_1^\textit{mmse} = \sqrt {{\beta}} {z^\textit{mmse}} = {\theta } + {\sigma _\textit{eff}}n$, the random vector ($\theta_k$,${\beta}_k$,$\widehat \theta_k^{\textit{mmse}}$) converges in distribution to the random vector ($\theta$,$\beta$,$\sqrt \beta  \widehat b$) as $N \to \infty $, which has the same distribution as that of the random vector ($\theta$,$\beta$,$\widehat \theta$). Hence, we have convergence in distribution to ($\theta$,$\beta$,$\widehat \theta$), and Property 2 follows from Property 1. For details on the proof of Property 1, we refer to \cite{DongningGuo2005}.
\end{IEEEproof}

Similar to Property 1, Property 2 shows that the MMSE estimation of the SMV problem in (3a) can be decoupled into scalar Gaussian noise corrupted linear problems as represented in (42a), and the corresponding noise variance in (42a) can be obtained by solving the fixed point equations in (11). Then it becomes obvious to find the MMSE estimates in (42b) and the corresponding mean square error.

Now, we seek to find the value of ${\sigma _{\textit{eff}}}$. Note that we can determine ${\sigma _{\textit{eff}}}$ by solving the fixed point equations in (\ref{equ3_7}). We refer to this method as $\textit{Property1}\text{-}{\sigma_{\textit{eff}}}$. The PMMSE estimator in (10b) is
\begin{equation}\label{mmse11}
\widehat b_{\textit{scalar}}^{\textit{ pmmse}}(z;{p_0}(b),{\eta _p}) = G\left( {|z{|^2};{\eta _p},\lambda ,1} \right)\frac{z}{{1 + {\eta _p}}}.
\end{equation}
Then, with (\ref{equ3_5}), the mean square error in (11b) is\cite{Tulino2013}
\begin{align}\label{mmse12}
&\textit{MSE}\left( {{p_0}(b),{p_0}(b),{\eta _p},{\eta _p},z} \right)  \nonumber\\
=& {\int {\left| {b - \widehat b_{\textit{scalar}}^{\textit{ pmmse}}(z;{p_0}(b),{\eta _p})} \right|} ^2}p\left( {b|z;{p_0}(b),{\eta _p}} \right)db  \nonumber\\
=& \lambda \left[ {1 - \frac{{\eta _p^2}}{{1 + {\eta _p}}}\omega \left( {\frac{{(1 + {\eta _p})(1 - \lambda )}}{{\lambda {\eta _p}}},{\eta _p}} \right)} \right]
\end{align}
where
\begin{equation*}
\omega (a,b) = \int_0^\infty  {\frac{{t{e^{ - bt}}}}{{1 + a{e^{ - t}}}}dt}.
\end{equation*}
Following the same procedure as in \cite{Tulino2013}, we can obtain the mean square error in (11a) as
\begin{align}\label{mmse13}
& \textit{MSE}\left( {{p_0}(b),{p_0}(b),{\eta _p},\eta ,z} \right)  \nonumber\\
=& {\int {\left| {b - \widehat b_{scalar}^{\textit{ pmmse}}(z;{p_0}(b),{\eta _p})} \right|} ^2}p\left( {b|z;{p_0}(b),\eta } \right)db  \nonumber\\
=& \lambda \left[ {1 - \frac{{2\eta _p^2(1 + {\eta _p})}}{{{{(1 + \eta )}^2}}}\omega \left( {\overline a ,\overline b } \right) + \frac{{\eta _p^2}}{{1 + \eta }}{\omega _2}(\overline a ,\overline b ,\overline c ,\overline d )} \right]
\end{align}
where
\begin{equation*}
{\omega _2}(a,b,c,d) = \int_0^\infty  {\frac{{t{e^{ - bt}}\left( {1 - c{e^{ - dt}}} \right)}}{{{{\left( {1 + a{e^{ - t}}} \right)}^2}}}dt},
\end{equation*}
\begin{equation*}
\overline a  = \frac{{(1 + {\eta _p})(1 - \lambda )}}{{\lambda {\eta _p}}},
\end{equation*}
\begin{equation*}
\overline b  = \frac{{{\eta _p}(1 + {\eta _p})}}{{1 + \eta }},
\end{equation*}
\begin{equation*}
\overline c  = \frac{{(1 + \eta )(1 - \lambda )}}{{\lambda \eta }},
\end{equation*}
\begin{equation*}
\overline d  = \overline b /\eta.
\end{equation*}
Then,  we determine ${\sigma _{\textit{eff}}}$ by substituting (\ref{mmse12}), (\ref{mmse13}) and the distribution of $\beta$ into the fixed point equations in (11).

With the CB-AMP algorithm and also the theoretical result in (42a), we have another approach to obtain ${\sigma _{\textit{eff}}}$, specifically by using the state evolution equation (\ref{equ3_21}). When the CB-AMP algorithm converges, the noise variance in (\ref{equ3_20}) should satisfy ${\xi ^{t + 1}} = {\xi ^t}$. Therefore, $\sigma _{\textit{eff}}^2$ satisfies the fixed point equation
\begin{equation}\label{mmse14}
\sigma _{\textit{eff}}^2 = \sigma _0^2 + \gamma E\left\{ {{{\left| {\theta  - {{\widehat \theta }_{\textit{mmse}}}} \right|}^2}} \right\}
\end{equation}
where ${\widehat \theta _{\textit{mmse}}}$ is given in (\ref{mmse8}). Therefore, we can obtain ${\sigma _{\textit{eff}}}$ by solving the fixed point equation (\ref{mmse14}). We call this method as $\textit{State}\text{-}{\sigma_{\textit{eff}}}$.

The mean square error of the MMSE estimate for the linear model (\ref{mmse6}) is
\begin{equation}\label{mmse15}
E\left\{ {{{\left| {\theta  - {{\widehat \theta }_{\textit{mmse}}}} \right|}^2}} \right\} = \lambda \left( {E\{ \beta \}-\int_{{\beta _{\min }}}^{{\beta _{\max }}} {\int_0^\infty  {f({\sigma _{\textit{eff}}},\beta ,t)dt} d\beta } } \right)
\end{equation}
where
\begin{equation*}
f({\sigma_{\textit{eff}}},\beta ,t) = \frac{{\lambda t p(\beta) {e^{ - t\sigma _{\textit{eff}}^2/\beta }}}}{{{{\left( {\frac{\beta }{{\sigma _{\textit{eff}}^2}} + 1} \right)}^2}\left( {\frac{\lambda }{{\beta  + \sigma _{\textit{eff}}^2}} + \frac{{(1 - \lambda ){e^{ - t}}}}{{\sigma _{\textit{eff}}^2}}} \right)}},
\end{equation*}
${\beta _{\min }} = {(2R)^{ - \alpha }}$, and ${\beta _{\max }} = d_0^{ - \alpha }$.

By now, we have introduced the $\textit{Property1}\text{-}{\sigma_{\textit{eff}}}$ and $\textit{State} \text{-} {\sigma _{\textit{eff}}}$ method to compute the effective noise level ${\sigma_\textit{eff}}$ in (42a). Although derived with the state evolution equations of the CB-AMP algorithm, the set up of $\textit{State} \text{-} {\sigma _{\textit{eff}}}$ method is conditioned on the theoretical result in (42a) which is introduced by the decoupling principle. In other words, decoupling principle in Property 2 provides the theoretical basis for the asymptotic analysis of SMV based MMSE estimation for sparse signal vectors with i.n.i.d. non-zero components in this paper, including both $\textit{Property1}\text{-}{\sigma_{\textit{eff}}}$ and $\textit{State} \text{-} {\sigma _{\textit{eff}}}$ method, and the likelihood ratio test introduced in the following sections.

Since the mean square error $E\{ {{{| {\theta  - {{\widehat \theta }_{\textit{mmse}}}} |}^2}}\}$ monotonically increases with ${\sigma _{\textit{eff}}}$ \cite{GuoWuShitzEtAl2011}, there exists at most one solution for the fixed point equation (46). However, the fixed point equations in (11) may have multiple solutions. Therefore, unlike in the $\textit{State} \text{-} {\sigma _{\textit{eff}}}$ method, we may need to search for the true solution among several values to find ${\sigma _{\textit{eff}}}$ when $\textit{Property1} \text{-} {\sigma _{\textit{eff}}}$ method is used. Thus, $\textit{State} \text{-} {\sigma _{\textit{eff}}}$ method leads to a more efficient numerical evaluation in computing ${\sigma _{\textit{eff}}}$, compared with the $\textit{Property1} \text{-} {\sigma _{\textit{eff}}}$ method, while the set up of the former is conditioned on the theoretical result which is introduced by the decoupling principle (facilitating the theoretical analysis that lead to the latter method).

\section{Likelihood Ratio Test With SMV Based MMSE Estimation}
In this section, we consider the activity detection problem based on the received signal at only one AP. From the previous section, we know that the MMSE estimation breaks down the received signal of the linear model (3a) into Gaussian noise corrupted scalar versions described in (42a). Then, we can obtain the corresponding likelihood functions as follows:
\begin{equation}\label{lik1}
p(z_1^{\textit{mmse}}|a = 1) = \frac{1}{{\pi \left( {\beta  + \sigma _{\textit{eff}}^2} \right)}}\exp \left( { - \frac{{{{\left| {z_1^{\textit{mmse}}} \right|}^2}}}{{\beta  + \sigma _{\textit{eff}}^2}}} \right),
\end{equation}
\begin{equation}\label{lik2}
p(z_1^{\textit{mmse}}|a = 0) = \frac{1}{{\pi \sigma _{\textit{eff}}^2}}\exp \left( { - \frac{{{{\left| {z_1^{\textit{mmse}}} \right|}^2}}}{{\sigma _{\textit{eff}}^2}}} \right).
\end{equation}
Now, the likelihood ratio is
\begin{align}\label{lik3}
R\left( {z_1^{\textit{mmse}}} \right) &= \frac{{p(z_1^{\textit{mmse}}|a = 1)}}{{p(z_1^{\textit{mmse}}|a = 0)}}  \nonumber\\
&  = \frac{{\sigma _{\textit{eff}}^2}}{{\beta  + \sigma _{\textit{eff}}^2}}\exp \left( {\frac{{\beta {{\left| {z_1^{\textit{mmse}}} \right|}^2}}}{{\sigma _{\textit{eff}}^2(\beta  + \sigma _{\textit{eff}}^2)}}} \right).
\end{align}
Finally, we can obtain the likelihood ratio test rule
\begin{equation}\label{lik4}
R\left( {z_1^{\textit{mmse}}} \right) \LRT{a=1}{a=0}\frac{{1 - \lambda }}{\lambda }.
\end{equation}
After some algebraic operations, we have the following threshold detection rule:
\begin{equation}\label{lik5}
{{{\left| {z_1^{\textit{mmse}}} \right|}^2}} \LRT{a=1}{a=0}l'\left( {\sigma _{\textit{eff}}^2,\lambda ,\beta } \right)
\end{equation}
where
\begin{equation*}
l'\left( {\sigma _{\textit{eff}}^2,\lambda ,\beta } \right) = \frac{{\sigma _{\textit{eff}}^2(\beta  + \sigma _{\textit{eff}}^2)}}{\beta }\log \left( {\frac{{(1 - \lambda )(\beta  + \sigma _{\textit{eff}}^2)}}{{\lambda \sigma _{\textit{eff}}^2}}} \right).
\end{equation*}
Then, the false alarm probability is
\begin{equation}\label{lik6}
P_F = E\left\{ {\exp \left( { - \frac{{l'\left( {\sigma _{\textit{eff}}^2,\lambda ,\beta } \right)}}{{\sigma _{\textit{eff}}^2}}} \right)} \right\}
\end{equation}
and the miss detection probability is
\begin{equation}\label{lik7}
P_M = 1 - E\left\{ {\exp \left( { - \frac{{l'\left( {\sigma _{\textit{eff}}^2,\lambda ,\beta } \right)}}{{\beta  + \sigma _{\textit{eff}}^2}}} \right)} \right\}
\end{equation}
where the expectation is over the large-scale fading coefficient $\beta$. Finally, the error probability is
\begin{equation}\label{lik71}
P_{\textit{err}} = (1 - \lambda )P_F + \lambda P_M.
\end{equation}

\section{Activity Detection With SMV Based MMSE Estimation in Cell-Free Massive MIMO Networks}
In this section, we consider activity detection in cell-free massive MIMO systems with massive connectivity based on the cooperation of received pilot signals at the entire set of APs. As noted before, with the CB-AMP algorithm, we can obtain the SMV based MMSE estimates of the effective channel coefficients from all users to every AP, and the received pilot signals at each AP are decomposed into scalar Gaussian noise corrupted versions which can be expressed as
\begin{equation}\label{mul1}
z_2^{\textit{mmse}} = {\theta _{\textit{ij}}} + {\sigma _{\textit{eff}}}n.
\end{equation}

Within this cooperative framework, there are two different approaches, namely, centralized and distributed activity detection. In the centralized detection method, received pilot signals at all APs are transmitted to the CPU, and the decisions on the activity of all users are made based on all received pilot signals. In the distributed detection method, each AP makes its own decisions on the activity of all users based on its own received pilot signals, and subsequently the decisions and the corresponding detection reliabilities are transmitted to the CPU. The final decisions on the activity of all users are made at CPU based on the decisions from all APs while taking into account corresponding detection reliabilities.
\subsection{Centralized Activity Detection}
We denote the decoupled Gaussian noise corrupted signal of the MMSE estimate of the effective channel coefficients from the $i$th user to all the APs as ${{\bf{z}}_i^{\textit{mmse}}} \in {{\cal C}^{1 \times M}}$, which can be expressed as
\begin{equation}\label{cen1}
{{\bf{z}}_i^{\textit{mmse}}} = {{\bm{\theta }}_i} + {\sigma _{\textit{eff}}}{\bf{n}}.
\end{equation}
As also done in \cite{ChenSohrabiYu2018} and \cite{LiuYu2018}, we assume that each component of the noise vector ${\bf{n}}$ is independent of other components. Note that the probability density function of ${\bm{\theta }}_i$ is
\begin{equation}\label{cen2}
p({{\bm{\theta }}_i}) = (1 - \lambda )\delta ({{\bm{\theta }}_i}) + \lambda {\cal C}{\cal N}({\bf{0}},{{\bf{\Lambda }}_i}),
\end{equation}
and the conditional probability density functions of ${{\bf{z}}_i^{\textit{mmse}}}$ are
\begin{equation}\label{cen3}
p\left( {{{\bf{z}}_i^{\textit{mmse}}}|a_i = 0} \right) = \frac{1}{{{\pi ^M}\sigma _{\textit{eff}}^{2M}}}\exp \left( { - \frac{{{{\left| {{{\bf{z}}_i^{\textit{mmse}}}} \right|}^2}}}{{\sigma _{\textit{eff}}^2}}} \right)
\end{equation}
and
\begin{equation}\label{cen4}
p\left( {{{\bf{z}}_i^{\textit{mmse}}}|a_i = 1} \right) = \frac{1}{{{\pi ^M}\left| {{{\bf{\Sigma }}_i}} \right|}}\exp \left( { - {{\bf{z}}_i^{\textit{mmse}}}{\bf{\Sigma }}_i^{ - 1}{{\left( {{{\bf{z}}_i^{\textit{mmse}}}} \right)}^H}} \right),
\end{equation}
where ${{\bf{\Sigma }}_i} = {{\bf{\Lambda }}_i} + \sigma _{\textit{eff}}^2{{\bf{I}}_M}$. Then we can obtain the probability density function
\begin{equation}\label{cen5}
p\left( {{{\bf{z}}_i^{\textit{mmse}}}} \right) = \left( {1 - \lambda } \right)p\left( {{{\bf{z}}_i^{\textit{mmse}}}|a_i = 0} \right) + \lambda p\left( {{{\bf{z}}_i^{\textit{mmse}}}|a_i = 1} \right).
\end{equation}
Note that the conditional probability density function is given by
\begin{equation}\label{cen6}
p({{\bf{z}}_i^{\textit{mmse}}}|{{\bm{\theta }}_i}) = \frac{1}{{{\pi ^M}\sigma _{\textit{eff}}^{2M}}}\exp \left( { - \frac{{{{\left| {{{\bf{z}}_i^{\textit{mmse}}} - {{\bm{\theta }}_i}} \right|}^2}}}{{\sigma _{\textit{eff}}^2}}} \right).
\end{equation}
Then we can obtain the posterior conditional probability density function as
\begin{equation}\label{cen7}
p({{\bm{\theta }}_i}|{{\bf{z}}_i^{\textit{mmse}}}) = \frac{{p({{\bf{z}}_i^{\textit{mmse}}}|{{\bm{\theta }}_i})p({{\bm{\theta }}_i})}}{{p\left( {{{\bf{z}}_i^{\textit{mmse}}}} \right)}}.
\end{equation}
Finally, we can express the MMSE estimate of ${{\bm{\theta }}_j}$ as
\begin{align}\label{cen8}
&\widehat {\bm{\theta }}_i^{\textit{mmse}} = E\left\{ {{{\bm{\theta }}_i}|{{\bf{z}}_i^{\textit{mmse}}}} \right\}  \nonumber\\
&  = \frac{{{{\bf{z}}_i^{\textit{mmse}}}{{\left( {\sigma _{\textit{eff}}^2{\bf{\Lambda }}_i^{ - 1} + {{\bf{I}}_M}} \right)}^{ - 1}}}}{{1 + \frac{{(1 - \lambda )\left| {{{\bf{\Sigma }}_i}} \right|}}{{\lambda \sigma _{\textit{eff}}^{2M}}}\exp \left( { - {{\bf{z}}_i^{\textit{mmse}}}\left( {\frac{1}{{\sigma _{\textit{eff}}^2}}{{\bf{I}}_M} - {\bf{\Sigma }}_i^{ - 1}} \right){{\left( {{{\bf{z}}_i^{\textit{mmse}}}} \right)}^H}} \right)}}.
\end{align}
It is obvious that the elements of $\widehat {\bm{\theta }}_i^{\textit{mmse}}$ are same with ${\widehat \theta _{\textit{mmse}}}$ obtained in (\ref{mmse8}) with the scalar Gaussian noise corrupted linear model. Since the channel coefficients are independent of each other, ${{{\bf{\Sigma }}_j}}$ is a diagonal matrix. Thus,
\begin{equation}\label{cen9}
\widehat {\bm{\theta }}_i^{\textit{mmse}} = \frac{{{{\bf{z}}_i^{\textit{mmse}}}{{\left( {\sigma _{\textit{eff}}^2{\bf{\Lambda }}_i^{ - 1} + {{\bf{I}}_M}} \right)}^{ - 1}}}}{{1 + \frac{{(1 - \lambda )}}{\lambda }\exp \left( { - M\left( {{\varsigma _i} - {\kappa _i}} \right)} \right)}}
\end{equation}
where
\begin{equation}\label{cen10}
{\varsigma _i} = \frac{{{{\bf{z}}_i^{\textit{mmse}}}\left( {{{\bf{I}}_M} - \sigma _{\textit{eff}}^2{\bf{\Sigma }}_i^{ - 1}} \right){{\left( {{{\bf{z}}_i^{\textit{mmse}}}} \right)}^H}}}{{M\sigma _{\textit{eff}}^2}}
\end{equation}
and
\begin{equation}\label{cen11}
{\kappa _i} = \frac{1}{M}\sum\nolimits_{j = 1}^M {\log \left( {{\beta _{ij}} + \sigma _{\textit{eff}}^2} \right)}  - 2\log \left( {{\sigma _{\textit{eff}}}} \right).
\end{equation}

As $M$ grows without bound, we have $\widehat {\bm{\theta }}_i^{\textit{mmse}} = {\bf{z}}_i^{\textit{mmse}}{\left( {\sigma _{\textit{eff}}^2{\bf{\Lambda }}_i^{ - 1} + {{\bf{I}}_M}} \right)^{ - 1}}$, which is MMSE estimate of ${{\bm{\theta }}_i}$ given ${\bf{z}}_i^{\textit{mmse}}$, when ${\varsigma _i} > {\kappa _i}$, and we have $\widehat {\bm{\theta }}_i^{\textit{mmse}}=0$ when ${\varsigma _i}<{\kappa _i}$. Therefore, we obtain the following threshold detection rule:
\begin{equation}\label{cen12}
\begin{cases}
a_i=1, \textit{if}{\kern 5pt} {\varsigma _i} > {\kappa _i}, \\
a_i=0, \textit{if}{\kern 5pt} {\varsigma _i} < {\kappa _i}.
\end{cases}
\end{equation}

%Now let's consider the theoretical performance of the threshold detection rule in (\ref{cen12}).
Let us denote $\widetilde {\bf{z}}_i^{\textit{mmse}} = {{\bf{z}}^{\textit{mmse}}}{\left( {{{\bf{I}}_M} - \sigma _{\textit{eff}}^2{\bf{\Sigma }}_i^{ - 1}} \right)^{1/2}}/{\sqrt M {\sigma _{\textit{eff}}}}$, $\varsigma _i^0 = {\varsigma _i}$ under the condition that $a_i=0$, and $\varsigma _i^1 = {\varsigma _i}$ under the condition that $a_i=1$. According to (\ref{cen1}), $\widetilde {\bf{z}}_i^{\textit{mmse}} \sim {\cal C}{\cal N}\left( {{\bf{0}},\frac{{{{\bf{I}}_M} - \sigma _{\textit{eff}}^2{\bf{\Sigma }}_i^{ - 1}}}{M}} \right)$ when $a_i=0$, and $\widetilde {\bf{z}}_i^{\textit{mmse}} \sim {\cal C}{\cal N}\left( {{\bf{0}},{{\bf{\Sigma }}_i}\left( {{{\bf{I}}_M} - \sigma _{\textit{eff}}^2{\bf{\Sigma }}_i^{ - 1}} \right)/\left( {M\sigma _{\textit{eff}}^2} \right)} \right)$ when $a_i=1$. Since ${\varsigma _i} =\widetilde {\bf{z}}_i^{\textit{mmse}}{\left( {\widetilde {\bf{z}}_i^{\textit{mmse}}} \right)^H}$ and ${{{\bf{\Sigma }}_i}}$ is a diagonal matrix, ${\varsigma _i}$ is the squared summation of independent and circularly symmetric complex Gaussian random variables.

Denote the $j$th element of $\widetilde {\bf{z}}_i^{\textit{mmse}}$ as $z_{ij}^{\textit{mmse}}$. Then, $z_{ij}^{\textit{mmse}} \sim {\cal C}{\cal N}\left( {0,\frac{{{\beta _{ij}}}}{{M\left( {{\beta _{ij}} + \sigma _{\textit{eff}}^2} \right)}}} \right)$ when $a_i=0$, and $z_{ij}^{\textit{mmse}} \sim {\cal C}{\cal N}\left( {0,\frac{{{\beta _{ij}}}}{{M\sigma _{\textit{eff}}^2}}} \right)$ when $a_i=1$. Therefore, ${(z_{ij}^{\textit{mmse}})^2} \sim \Gamma \left( {1,\frac{{{\beta _{ij}}}}{{M\left( {{\beta _{ij}} + \sigma _{\textit{eff}}^2} \right)}}} \right)$ when $a_i=0$, and ${(z_{ij}^{\textit{mmse}})^2} \sim \Gamma \left( {1,\frac{{{\beta _{ij}}}}{{M\sigma _{\textit{eff}}^2}}} \right)$ when $a_i=1$, where $\Gamma (a,b)$ stands for the Gamma distribution with shape parameter $a$ and scalar parameter $b$. We denote $\rho _{ij}^0 = \frac{{{\beta _{ij}}}}{{M\left( {{\beta _{ij}} + \sigma _{\textit{eff}}^2} \right)}}$ and $\rho _{ij}^1 = \frac{{{\beta _{ij}}}}{{M\sigma _{\textit{eff}}^2}}$.

Since $\frac{x}{{1 + x}} < \log (1 + x) < x$ when $x>0$, we can obtain
\begin{equation}\label{cen17}
\sum\nolimits_{j = 1}^M {\rho _{ij}^0}  < {\kappa _i} < \sum\nolimits_{j = 1}^M {\rho _{ij}^1}.
\end{equation}
With the law of large numbers, we have ${\varsigma _i^0} \to \sum\nolimits_{j = 1}^M {\rho _{ij}^0}$ and ${\varsigma _i^1} \to \sum\nolimits_{j = 1}^M {\rho _{ij}^1}$ as $M$ grows. Therefore, ${\varsigma _i^0} < {\kappa _i}$ and ${\varsigma _i^1}> {\kappa _i}$ when $M \to \infty $. Thus, the false alarm and miss detection probabilities of the threshold detection rule in (\ref{cen12}) tend to zero as $M \to \infty $. So does the error probability.

\textbf{Remark 5:} In order to compare the performances of different centralized activity detection methods, error probability of centralized activity detection with MMV based MMSE estimation in cell-free massive MIMO networks (which has been studied in \cite{Guo-ISIT20}) is also provided in Section \uppercase\expandafter{\romannumeral8} via numerical results.

\subsection{Distributed Activity Detection}
For distributed activity detection, each AP acquires the MMSE estimates of the effective channel coefficients from all users to this AP with its own received pilot signals. Subsequently, each AP makes its own decisions on the activity of all users based on the MMSE estimates and the threshold detection rule in (51), and sends these decisions and the corresponding reliabilities to the CPU. Then the final decisions on the activity of all users are made at the CPU based on the decisions from the entire set of APs, while also taking into account the corresponding reliabilities. The detection reliability of the decisions made at each AP can be determined via (\ref{lik6}) and (\ref{lik7}).

Let us denote the decision made by the $j$th AP for the $i$th user as $d_{ij}$, and consider the optimal fusion rule for the activity of the $i$th user. Based on $d_{ij}$ ($j = 1,2, \cdots M$), we divide all the APs into two sets, ${{\cal S}_0} = \left\{ {j|{d_{ij}} = 0} \right\}$ and ${{\cal S}_1} = \left\{ {j|{d_{ij}} = 1} \right\}$. As shown in \cite{Chair1986}, the optimal fusion rule is
\begin{equation}\label{dis1}
 \left| {{{\cal S}_1}} \right| {\log \left( {\frac{{1 - P_{M}}}{P_{F}}} \right)}  + \left| {{{\cal S}_0}} \right| {\log \left( {\frac{P_{M}}{{1 - P_{F}}}} \right)}\LRT{{{a_i} = 1}}{{{a_i} = 0}} \log \left( {\frac{1-\lambda }{{\lambda }}} \right),
\end{equation}
where ${P_{F}}$ and ${P_{M}}$ are given in (\ref{lik6}) and (\ref{lik7}), respectively. After several algebraic operation, it is easy to obtain the following fusion rule:
\begin{equation}\label{dis2}
\left| {{{\cal S}_1}} \right|\log \frac{{(1 - P_M)(1 - P_F)}}{{P_M P_F}} \LRT{a_i=1}{a_i=0}\log \frac{{1 - \lambda }}{\lambda } - M\log \frac{{P_M}}{{1 - P_F}}.
\end{equation}

Let us set ${\chi} = \log \frac{{( {1 - P_M})( {1 - P_F} )}}{{P_M P_F}}$ and ${\rho} = ( {\log \frac{{1 - \lambda }}{\lambda } - M\log \frac{{P_M}}{{1 - P_F}}} )/{\chi}$. We assume $\lambda  < 0.5$ in this paper\footnote{Massive connectivity and small active probability for each user are two important characteristics of IoT networks.}. Then $l'\left( {\sigma _{\textit{eff}}^2,\lambda ,\beta } \right) > 0$ and
\begin{equation}\label{dis2_1}
\exp \left( { - \frac{{l'\left( {\sigma _{\textit{eff}}^2,\lambda ,\beta } \right)}}{{\sigma _{\textit{eff}}^2}}} \right) < \exp \left( { - \frac{{l'\left( {\sigma _{\textit{eff}}^2,\lambda ,\beta } \right)}}{{\beta  + \sigma _{\textit{eff}}^2}}} \right).
\end{equation}
Therefore,
\begin{equation}\label{dis2_2}
P_F+ P_M< 1.
\end{equation}
Thus, ${\chi} > 0$. Then, the optimal fusion rule is
\begin{equation}\label{dis3}
\left| {{{\cal S}_1}} \right| \LRT{a_i=1}{a_i=0}{\rho}.
\end{equation}

When $a_i=0$, we have $\left| {{{\cal S}_1}} \right| \sim B(M,P_F)$, where $B(M,P_F)$ stands for the binomial distribution. Then we can obtain the false alarm probability as
\begin{equation}\label{dis5}
P_F^d =1 - \psi \left( {\left\lfloor {{\rho}} \right\rfloor } \right)
\end{equation}
where $\psi \left(  \cdot  \right)$ stands for the cumulative distribution function of the binomial distribution, and $\left\lfloor x \right\rfloor $ is the largest integer which is smaller than or equal to $x$. Similarly, when $a_i=1$, $\left| {{{\cal S}_1}} \right| \sim B(M,1-P_M)$, and the corresponding miss detection probability of the system is
\begin{equation}\label{dis6}
P_M^d =\psi \left( {\left\lfloor {{\rho}} \right\rfloor } \right).
\end{equation}
Finally, the error probability of the system is
\begin{equation}\label{dis7}
P_{\textit{err}}^d = (1 - \lambda )P_F^d + \lambda P_M^d.
\end{equation}

%\textbf{Remark:} Although we can obtain the false alarm, miss detection and error probabilities with (\ref{dis5}), (\ref{dis6}) and (\ref{dis7}) directly from the CDF of corresponding binomial distributions, the cumulative calculating error increases as $M$ increasing when we do the numerical simulations. Especially when $M$ is large while $P_F$ or $1-P_M$ is very small. On the other hand, $\left| {{{\cal S}_1}} \right|$ can be regarded as the summation of $M$ independent and identically distributed Bernoulli random variables. It is easy to verify that the assumptions in theorem 1 are satisfied, and we can use theorem 1 to calculate the false alarm, miss detection and error probabilities in (\ref{dis5}), (\ref{dis6}) and (\ref{dis7}). Numerical results in Section \uppercase\expandafter{\romannumeral7} show that the results obtained with theorem 1 is much more accurate than that achieved with (\ref{dis5}), (\ref{dis6}) and (\ref{dis7}), when $M$ is large.

We define $f(x) = P_F\log (x) + (1 - P_F)\log (1 - x)$, $0 < x < 1$. It can be readily shown that $f(x)$ is monotonically decreasing within the interval $(P_F,1]$. Since $P_F + P_M< 1$, we have $f(P_F) > f(1 - P_M)$, i.e.,
\begin{align}\label{dis8}
&P_F\log (P_F) + (1 - P_F)\log (1 - P_F) > \nonumber\\
&{\kern 60pt}P_F\log (1 - P_M) + (1 - P_F)\log (P_M).
\end{align}
Thus,
\begin{align}\label{dis9}
&P_F\log \left( {\frac{{1 - P_M}}{{P_F}}} \right) < \left( {1 - P_F} \right)\log \left( {\frac{{1 - P_F}}{{P_M}}} \right) \nonumber\\
&{\kern 140pt}+ \frac{1}{M}\log \left( {\frac{{1 - \lambda }}{\lambda }} \right).
\end{align}
Then, we can obtain
\begin{equation}\label{dis10}
\mathop {\lim }\limits_{M \to \infty } \frac{{{\rho }}}{{MP_F}} > 1.
\end{equation}
Similarly, we can also obtain
\begin{equation}\label{dis11}
\mathop {\lim }\limits_{M \to \infty } \frac{{{\rho}}}{{M\left( {1 - P_M} \right)}} < 1.
\end{equation}

As $M$ grows, we have $\left| {{{\cal S}_1}} \right| = MP_F$ when $a_i=0$ and $\left| {{{\cal S}_1}} \right| = M\left( {1 - P_M} \right)$ when $a_i=1$. With (\ref{dis3}), (\ref{dis10}) and (\ref{dis11}), we note that the false alarm and miss detection probabilities of the optimal fusion rule tend to zero as $M \to \infty $. Thus, the error probability of the optimal fusion rule also tends to zero when $M$ grows without bound.

In cell-free massive MIMO systems, the backhaul link is from the APs to the CPU since there is no information exchange between APs. As a consequence of transmitting only activity decisions and the corresponding reliabilities by each AP, the capacity of backhaul link which is needed for distributed activity detection is much smaller than that needed for centralized activity detection, which sends the entire received pilots to the CPU by every AP.

\textbf{Remark 6:} In this section, we have arrived at the conclusion that the error probabilities of both centralized and distributed activity detection tend to zero when the number of APs grows without bound. Since cell-free massive MIMO is a more general model that can be specialized to a single-cell massive MIMO system when all APs are located at the same node, these results confirm and generalize the conclusion in \cite{LiuYu2018} that the error probabilities of activity detection in a single-cell massive MIMO system tend to zero when the number of antennas at base station goes to infinity.

\section{Extensions to MMV Based MMSE Estimation and Activity Detection}

Heretofore, we have considered the theoretical analysis of SMV based MMSE estimation for effective channel coefficients in cell-free massive MIMO systems with massive connectivity, and the corresponding theoretical analysis on error probabilities of activity detection rules with SMV based MMSE estimation results. In Section \uppercase\expandafter{\romannumeral4}, the theoretical mean square error of SMV based MMSE estimation can be obtained with the $\textit{Property1} \text{-} {\sigma _{\textit{eff}}}$ or $\textit{State} \text{-} {\sigma _{\textit{eff}}}$ method and equation (46). The theoretical error probability of likelihood ratio test with SMV based MMSE estimation results depending on the received pilot signals at only one AP is shown in equations (52), (53) and (54) of Section \uppercase\expandafter{\romannumeral5}. Section \uppercase\expandafter{\romannumeral6} addresses the theoretical analysis of likelihood ratio test with SMV based MMSE estimation results and cooperation among different APs. We do not obtain closed-form theoretical error probabilities in this section, but these probabilities are in general dependent on $\sigma _{\textit{eff}}$, and they can be determined via numerical analysis, i.e., by substituting the theoretical value of $\sigma _{\textit{eff}}$ into (56) and finding the corresponding error probability numerically. Moreover, with the theoretical analysis, we find that the error probabilities of both centralized and distributed activity detection tend to zero when the number of APs tends to infinity while the asymptotic ratio between the number of users and pilots is kept constant.

In order to simplify theoretical analysis, we consider only SMV based methods in Section \uppercase\expandafter{\romannumeral4} through Section \uppercase\expandafter{\romannumeral6}. However, since equation (\ref{equ4}) presents an MMV problem, there can be a performance gap when the correlation between different columns of $\bf{Y}$ are not taken into account, and (\ref{equ4}) is decomposed into multiple SMV problems. On the other hand, it is challenging to perform a theoretical analysis of MMV based MMSE estimation for effective channel coefficients of cell-free massive MIMO systems with massive connectivity. Therefore, the performance loss between SMV and MMV based methods is shown via numerical results. In the next section, AMP algorithm for MMV proposed in \cite{ChenSohrabiYu2018} is employed to obtain several numerical results for the mean square error of MMV based MMSE estimation, and the error probability of centralized activity detection with MMV based MMSE estimation results are also determined numerically.

\section{Numerical Results}
\begin{figure}[htbp]
	\center
	\includegraphics[width=3.5in]{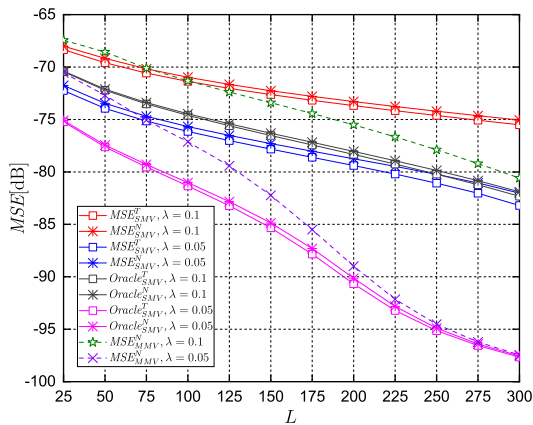}\\
	\caption{Mean square error of the oracle, SMV and MMV based MMSE estimation in cell-free Massive MIMO networks with massive connectivity versus the number of pilots $L$.}\label{fig1}
\end{figure}

\begin{figure}[htbp]
	\center
	\includegraphics[width=3.5in]{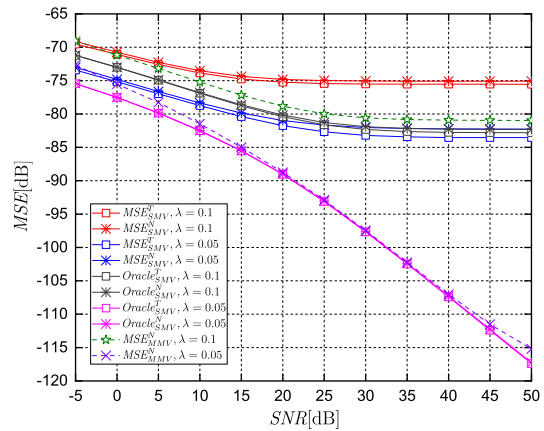}\\
	\caption{Mean square error of the oracle, SMV and MMV based MMSE estimation in cell-free Massive MIMO networks with massive connectivity versus SNR at each AP.}\label{fig2}
\end{figure}

\begin{figure}[htbp]
	\center
	\includegraphics[width=3.5in]{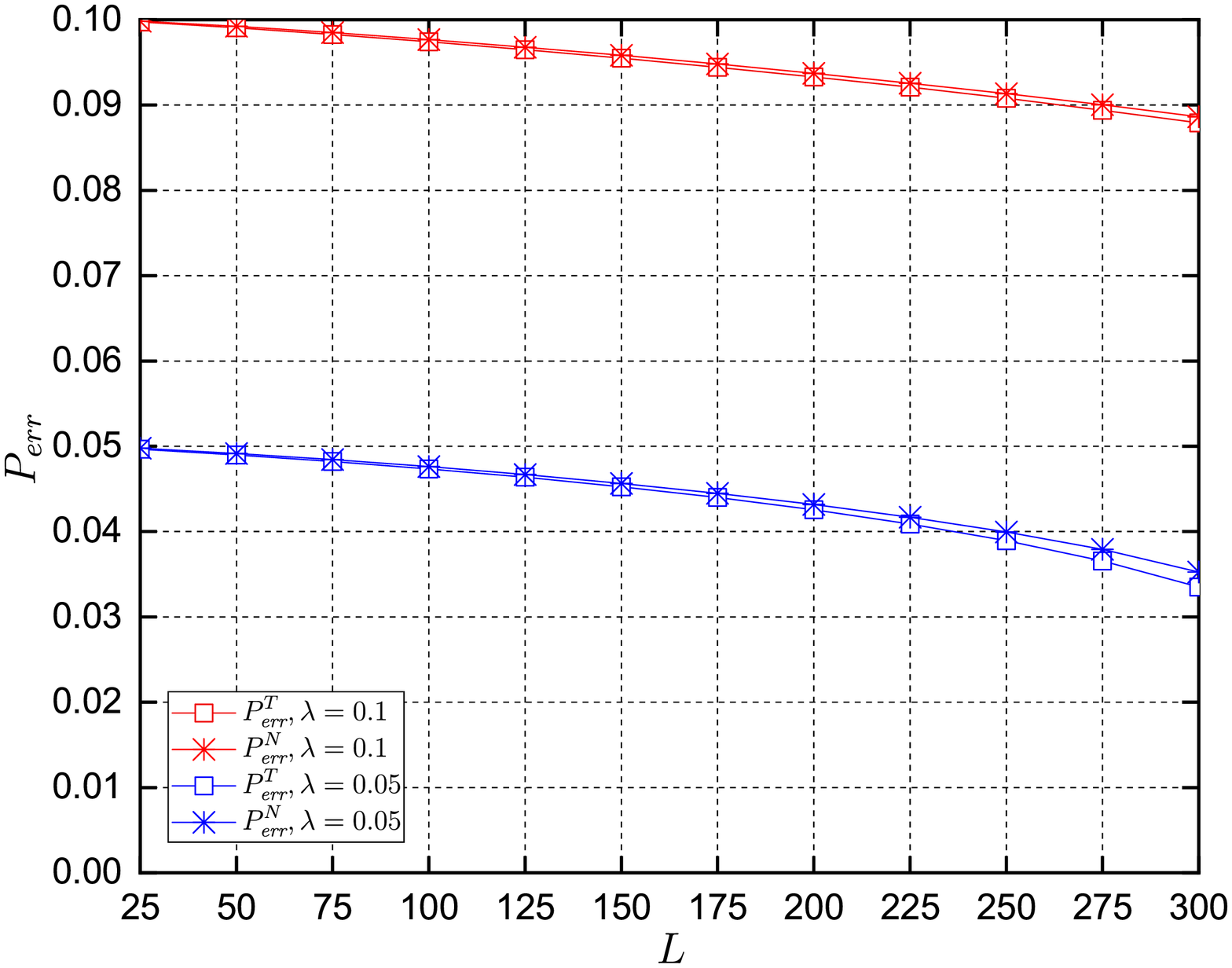}\\
	\caption{Error probability of likelihood ratio test based on the received pilot signals at only one AP versus the number of pilots used during the MMSE estimation.}\label{fig3}
\end{figure}

\begin{figure}[htbp]
	\center
	\includegraphics[width=3.5in]{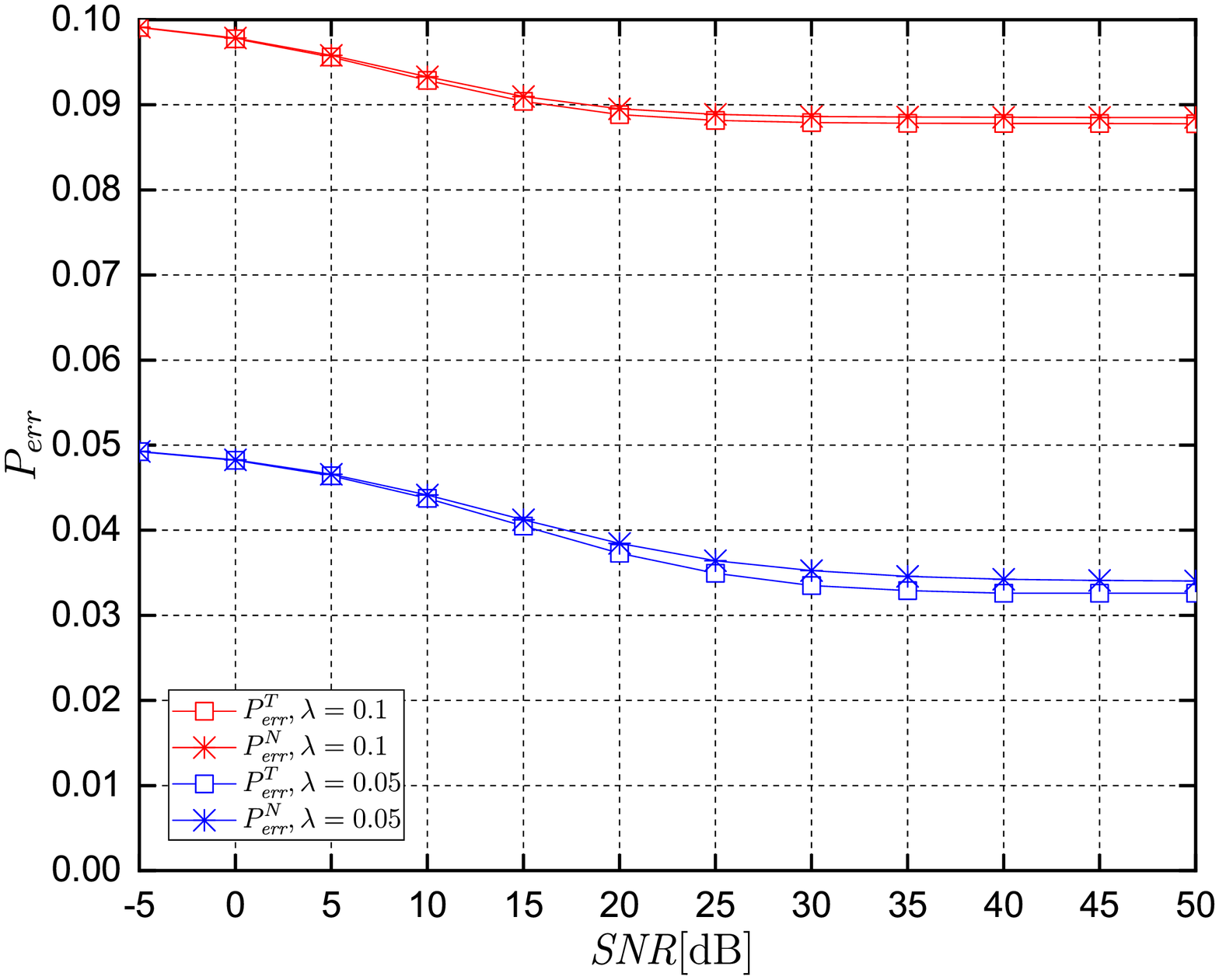}\\
	\caption{Error probability of likelihood ratio test based on the received pilot signals at only one AP versus SNR.}\label{fig4}
\end{figure}

\begin{figure}[htbp]
	\center
	\includegraphics[width=3.5in]{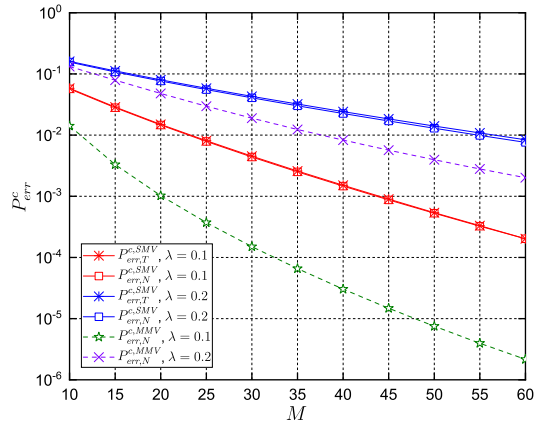}\\
	\caption{Error probability of centralized detection in cell-free Massive MIMO networks with massive connectivity versus the number of APs.}\label{fig5}
\end{figure}

\begin{figure}[htbp]
	\center
	\includegraphics[width=3.5in]{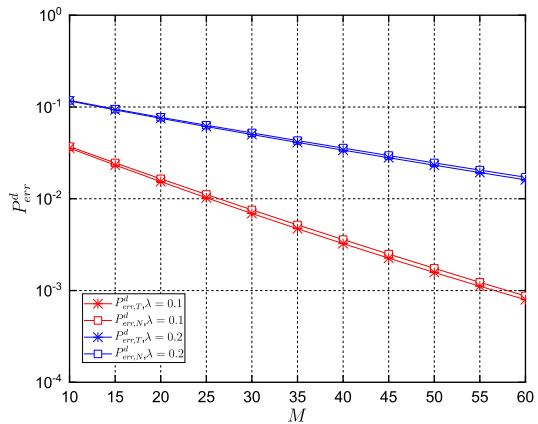}\\
	\caption{Error probability of distributed detection with the optimal fusion rule in cell-free massive MIMO networks with massive connectivity versus the number of APs.}\label{fig6}
\end{figure}

In the setting for numerical analyze, we assume that the APs and users are uniformly distributed in a circular area with radius $R=500$m. The path loss decay exponent is $\alpha=2.5$, and the reference distance is set at $d_0=50$m. The total user number is $N=4000$. CB-AMP and AMP for MMV algorithms are employed to obtain numerical results for the mean square error of SMV and MMV based MMSE estimates, respectively, of the effective channel coefficients in cell-free massive MIMO systems with massive connectivity. In the simulation settings of Figs. \ref{fig1} and \ref{fig3}, we consider 10 APs in cell-free massive MIMO system, and the SNR at each AP is 30dB. In Figs. \ref{fig2} and \ref{fig4}, we again consider 10 APs and assume that the number of pilots used for MMSE estimation is 300. In Figs. \ref{fig5} and \ref{fig6}, the number of pilots used for MMSE estimation is 300, and the SNR at each AP is 30dB.

In the numerical results, we analyze the performance of MMSE estimation and activity detection considering different probabilities of each user being active. We plot the curves of the mean square error of oracle and MMSE estimators as a function of the number of pilots used for estimation and the SNR at each AP. The error probabilities of the likelihood ratio test based on MMSE estimation is also analyzed in this section. Additionally, we study the error probabilities of centralized and distributed detection in cell-free massive MIMO systems versus the number of APs.

For the simplification of descriptions, we first introduce several abbreviations that will be used in the figures of the numerical results. ``${\textit{MSE}}_\textit{SMV}^T$'' and ``${\textit{MSE}}_\textit{SMV}^N$'' stand for the theoretical and numerical mean square errors of the MMSE estimation obtained with the replica analysis result (\ref{mmse15}) and CB-AMP algorithm, respectively. ``${\textit{MSE}}_\text{MMV}^N$'' denotes the mean square error of MMSE estimation acquired with the AMP for MMV algorithm.  ``${\textit{Oracle}}_\textit{SMV}^T$'' and ``${\textit{Oracle}}_\textit{SMV}^N$'' indicate the theoretical and numerical mean square errors of the oracle estimation obtained with the asymptotic result (\ref{asym9}) and the numerical result (\ref{oracle5}), respectively. ``$P_\textit{err}^T$'' and ``$P_\textit{err}^N$'' denote the theoretical and numerical error probabilities of the likelihood ratio test based on the decoupled scalar model (\ref{mmse6}) and the AWGN corrupted scalar model (\ref{equ3_20}) involved with the CB-AMP algorithm, respectively. Similarly, ``$P^{\textit{c,SMV}}_\textit{err,T}$'' and ``$P^{\textit{c,SMV}}_\text{err,N}$'' stand for the theoretical and numerical error probabilities for the centralized likelihood ratio test with SMV based MMSE estimation results in cell-free massive MIMO systems. ``$P^d_\textit{err,T}$'' and ``$P^d_\text{err,N}$'' designate the theoretical and numerical results of error probabilities of the distributed likelihood ratio test under the optimal fusion rule (\ref{dis3}) in cell-free massive MIMO systems. ``$P^{\textit{c,SMV}}_\textit{err,T}$'' and ``$P^d_\textit{err,T}$'' are evaluated with the scalar Gaussian noise corrupted linear model (\ref{mul1}), while ``$P^{\textit{c,SMV}}_\textit{err,N}$'' and  ``$P^d_\text{err,N}$'' are obtained with the CB-AMP algorithm and its state evolution equations (\ref{equ3_20}). ``$P^{\textit{c,MMV}}_\textit{err,N}$'' denotes the error probability of centralized activity detection based on the MMSE estimates of the effective channel coefficients via AMP for MMV algorithm and the corresponding state evolution equations.

Fig. \ref{fig1} plots the mean square error of the oracle, SMV and MMV based MMSE estimation in cell-free massive MIMO networks with massive connectivity as a function of the number of pilots $L$. These curves show that the numerical and theoretical results of oracle and SMV based MMSE estimation match well with each other. The mean square error decreases as the number of pilots increases, and the rate of decay in the curves increases as the probability of user being active decreases. Furthermore, compared with the MMV based MMSE estimation, the decay rate of the mean square error of SMV based estimation is smaller. The former has larger mean square error when $L$ is small while it has smaller values when $L$ is larger than 75. Moreover, we observe that as a consequence of the uncertainty in users' activities, compared with the oracle estimation, under the same settings, the mean square error of SMV based MMSE estimation is larger, and the gap between them increases as $\lambda$ decreases from 0.1 to 0.05.

Fig. \ref{fig2} plots the mean square error of oracle, SMV and MMV based MMSE estimation in cell-free Massive MIMO networks with massive connectivity versus SNR at each AP. When SNR is small, the mean square errors diminish as SNR increases. However, when SNR is large enough, the tendency of the mean square error becomes different. Since the total number of users is 4000, the expected number of active users is 400 when $\lambda=0.1$, and this becomes 200 when $\lambda=0.05$. Recall that the number of pilots used for estimation is 300. Therefore, when $\lambda=0.1$, the received signal which is comprised of the transmitted signal from all active users constitutes an underdetermined linear system. Thus, the mean square error of the oracle estimation becomes flat when SNR becomes large enough. As a consequence of the system being underdetermined and the uncertainty of users' activities, the mean square errors of both SMV and MMV based MMSE estimation also flatten out when SNR becomes large enough. On the contrary, the received signal from all active users constitutes an overdetermined linear system when $\lambda=0.05$. Therefore, the mean square error of oracle estimation keeps decreasing as SNR increases, and tends to zero as SNR grows without bound. However, since users' activities are unknown to the APs, the mean square error of both SMV and MMV based MMSE estimation still flatten out after SNR reaches a certain high level when $\lambda=0.05$. Moreover, we have also noticed that the theoretical mean square error of oracle and SMV based MMSE estimation match well with that of the numerical results. These curves also show that the mean square error of the MMV based MMSE estimation is always smaller than that of SMV based results under the same settings.

Fig. \ref{fig3} plots the theoretical and numerical results for the error probability of the likelihood ratio test based on the received pilot signals at only one AP versus the number of pilots used in the MMSE estimation. We see that the error probabilities diminish as the number of pilots increases. We also note that when $\lambda$ decreases, the error probabilities diminish more quickly.

Fig. \ref{fig4} plots the theoretical and numerical results for the error probabilities of the likelihood ratio test based on the received pilot signals at only one AP versus SNR. From the analysis of Fig. \ref{fig2}, we know that the mean square error of MMSE estimation diminishes as SNR initially grows starting from small values. After SNR becomes large enough, the mean square error flattens out. Since the likelihood ratio test is based on the MMSE estimation result, as SNR increases, error probabilities should also decrease with SNR initially, and then become flat when the SNR is beyond a certain threshold. Comparing Figs. \ref{fig2} and \ref{fig4}, we can see that the mean square errors and error probabilities becomes flat at the same SNR value within the same numerical simulation settings, which is consistent with our theoretical analysis.

Figs. \ref{fig5} and \ref{fig6} plot the Error probabilities of centralized and distributed detection in cell-free massive MIMO networks with massive connectivity versus the number of APs, respectively. The error probabilities diminish as the number of APs increases, confirming our conclusion that the error probabilities of both centralized and distributed detection tend to zero when the number of AP tends to infinity while the ratio between the number of users and pilots is kept constant. When the user activity probability decreases from 0.2 to 0.1, the error probabilities decrease, and the rate of decrease becomes larger. Furthermore, in Fig. \ref{fig5}, we plot the curves for centralized detection with both SMV and MMV based MMSE estimation results of the effective channel coefficients. The gaps between the SMV and MMV based results show the performance loss due to disregarding the correlation between different columns of the received pilot signals when performing channel estimation.

\section{Conclusion}
In this paper, we have derived a decoupling principle for SMV based MMSE estimation of sparse signal vectors with i.n.i.d. non-zero components in cell-free massive MIMO networks with massive connectivity. With this decoupling principle and the likelihood ratio test, we have obtained a detection rule for the activity of users based on the received pilot signals at only one AP. Subsequently, with the decoupling principle of the SMV based MMSE estimation, likelihood ratio test and the optimal fusion rule, we have determined detection rules for the activity of users based on the cooperation with the received pilot signals at all APs for centralized and distributed detection. We have also demonstrated that the error probabilities of both centralized and distributed detection tend to zero when the number of AP tends to infinity while the asymptotic ratio between the number of users and pilots is kept constant. Moreover, we have analyzed oracle estimation in cell-free massive MIMO networks with massive connectivity, and the asymptotic behavior of oracle estimation is identified via random matrix theory. Via numerical analysis, we have investigated the impact of the number of pilots, SNR, and the number of APs on mean square error and error probabilities of different schemes.  We have also observed that the theoretical analysis with the decoupling principle of SMV based MMSE estimation matches well with the numerical results of the CB-AMP algorithm.

%\appendices

\bibliographystyle{IEEEtran}
\bibliography{asymptotic}

\end{document}